\documentclass[review]{elsarticle}
\usepackage{graphicx}
\usepackage{lineno,hyperref,}
\usepackage[cmex10]{amsmath}
\usepackage[margin = 1.0in]{geometry}
\modulolinenumbers[5]

\usepackage{amsmath}
\usepackage{floatrow}
\usepackage{eurosym}
\usepackage{subfigure}

\DeclareMathOperator{\dd}{\mathrm{d} \!}
\DeclareMathOperator{\pr}{\text{pr}}

\begin{document}

\begin{frontmatter}

\title{Minority games played by arbitrageurs on the energy market}

\author{Tim Ritmeester and Hildegard Meyer-Ortmanns}
\address{Physics and Earth Sciences\\
	Jacobs University Bremen\\
	28759 Bremen, Germany\\
	Email: h.ortmanns@jacobs-university.de}

\begin{abstract}
Along with the energy transition, the energy markets change their organization toward more decentralized  and self-organized structures, striving for locally optimal profits. These tendencies may endanger the physical grid stability. One realistic option is the exhaustion of reserve energy due to an abuse by arbitrageurs. We map the energy market to different versions of a minority game and determine the expected amount of arbitrage as well as its fluctuations  as a function of the model parameters. Of particular interest are the impact of heterogeneous contributions of arbitrageurs, the interplay between external stochastic events and nonlinear price functions of reserve power, and the effect of risk aversion due to suspected penalties. The non-monotonic dependence of arbitrage on the control parameters  reveals an underlying phase transition that is the counterpart  to replica symmetry breaking in spin glasses. As conclusions from our results we propose economic and statutory measures to counteract a detrimental effect of arbitrage.
\end{abstract}


\end{frontmatter}


\section{Introduction}\label{sec: introduction}
Along with the energy transition, not only the physical realization of the power grid gets reorganized towards a more decentralized structure, but also  energy markets try to act more self-organized and on a spatially local platform\footnote{local in contrast to country-wide}, striving for their own  optimum of profit. Occasionally, these efforts for making maximal profit in a given local trading area of the grid may endanger the global physical grid stability.
Supply and demand of power need to be  balanced to ensure the stability of the power transmission grid. Any entity that wishes to trade on the European energy market is obligated to be managed by a Balancing Responsible Party (BRP). It has a legal obligation to match the supply and demand of power in their portfolio to the best of their abilities \cite{50_hertz_untersuchung_2019}. However, deviations from this delicate balance are inevitable due to unforeseen contingencies, fluctuating renewable power sources from wind and solar energy, and fluctuations in the power consumption. Responsible for the stability of the power grid itself are the Transmission System Operators (TSOs), who counteract these deviations by activating reserve power, which is kept in reserve. A TSO is an entity entrusted with transporting energy in the form here of electrical power on a national or regional level, based on a fixed infrastructure. The term is defined by the European Commission.
The reserve power can be both positive  (in case of lack of supply) or negative (in case of lack of demand). It has some cost, which is paid for by BRPs  proportionally to the imbalance on their portfolios. Of course there is only a limited amount of reserve necessary.

Reserve power is offered at an auction, separately from other energy markets. Offers correspond to a promise to keep a certain amount of power in reserve for a given 4-hour interval, and to activate the power when requested.
Apart from this auction at the reserve power market, our discussion will focus on the so-called intraday or spot market. Trade on the intraday market concerns power to be delivered within a given 15-min interval, on the same day as on which it is traded.

Since suppliers of reserve energy have to obey relatively strict requirements (e.g. the ability to activate the power within minutes), the cost of reserve power in general exceeds the cost of power purchased on the regular energy markets such as the intraday market. Hence the BRPs have, besides a legal incentive, also an economic incentive to balance their positions.
Nevertheless it may happen that the reserve energy price falls below the  energy price on other markets, here specifically below the price on the intraday market.

Offers for reserve power have to be made a day in advance, while trade on the intraday market can take place up to 15 minutes before the delivery itself. Changes in circumstances between these time-frames, e.g. altered weather predictions, can therefore change the price on the intraday market in such a way that it is higher than the price of reserve power. This gives rise to a situation in which it is economically advantageous for a BRP to create a virtual imbalance in his portfolio: first short selling power on the intraday market and afterwards effectively buying reserve power at a lower price by forcing the TSO to activate reserve power for compensation. Such an instance corresponds to arbitrage: The BRP makes profit from the price difference apparently without any risk.

In economics and finance, arbitrage is the practice of taking advantage of a price difference between two or more markets without risk, the profit being the difference between the market prices at which the unit (here energy) is traded. From a physics background it is interesting to note that arbitrage in economics realizes the concept of frustration \cite{mack1}. In abstract terms, what is in common is the procedure that a quantity  is ``parallel transported'' around a closed loop and changes upon this transport \footnote{The concept of frustration is familiar from spin glasses, oscillatory systems, biological applications, power grids and others \cite{hmopablo}. In general, the dynamics of a system is considerably enriched if frustration is inherent in the system due to conflicting input, where the ``conflict is solved'' by a proliferation of stationary states  such as the ground states in spin glasses.}. Here it is the energy which is bought and sold (possibly a number of times) by the same BRP such that the process picks up a price difference in the end.

In the context of the energy market, arbitrage should be avoided as it may destabilize the grid: It may happen that more reserve energy is required than is accessible. However, such arbitrage opportunities have been actually exploited on the German energy market at several occasions in June 2019; this behavior has almost led to a crash of the system as not enough reserve power was immediately at disposal. We will use data of these events later in our discussion.

In reality, different types of reserve power are distinguished. Relevant for us are secondary and tertiary reserve power, of which a total of  around $2500$ MW and $1500 $ MW are available in Germany and Austria, respectively. Both have different start-up times, and whether (and when) tertiary reserve power gets activated depends on the expected duration of the disturbance. In our model we consider a single reserve market, which represents the combination of these two types of reserve power. The price of the reserve power  is determined by the so-called merit order. Merit order means that the cheapest power is activated first, and the larger the imbalance is, the larger is the amount of reserve power that is activated at a higher price. The merit order is essential to prevent an arbitrage procedure from becoming  self-supporting, ending in a runaway process. From the point of view of the BRPs, the merit order implies a limitation of  arbitrage opportunities: The more BRPs engage in this behaviour, the higher the cost of the reserve power, until eventually the possibility for arbitrage disappears.

Even with the merit order in place, this leaves room for some amount of arbitrage, thereby using precious reserve power. Here, the incentives for the BRPs amount to a so-called 'minority mechanism': Performing arbitrage can be only advantageous for a BRP if not many other BRPs  behave in the same way. How many behave in the same way is not known in advance, since any BRP only gets to find out what the reserve energy price is  \textit{after} he makes his decision (the trade on the intraday market stops 15 minutes before the actual delivery of the power).  Thus the minority mechanism leads to an uncertainty in who actually contributes to buying reserve power and in the reserve power price. This leads to the possibility that some BRPs overestimate the profit they can make by leaving their portfolio imbalanced by means of short selling,  favouring arbitrage by ignoring  possibly similar actions of other BRPs. The overestimation of the possible profit leads to an increased risk of exhausting the reserve power.

Thus, from the point of view of the BRPs, they should try to \textit{anti-coordinate} to the average behaviour: estimate the total arbitrage performed by the other BRPs, and anti-align their own behaviour with the aggregated behaviour of the other BRPs (that is,   refrain from getting involved in arbitrage if many others get involved, otherwise perform arbitrage). Therefore, the BRPs, our agents, have to learn the behaviour of the other agents,  and adjust their own behaviour accordingly. This is not straightforward, as the 'best strategy' must not be the same for all: If all  agents would use the same strategy, they would come to the same decision, and the strategy would invalidate itself. Anti-coordination thus requires the agents to reach some degree of heterogeneity in making their decisions. In case they would manage to perfectly anti-coordinate and always estimate the reserve price correctly,  there were no fluctuations around the average amount of arbitrage and no additional risk caused by fluctuations of their actions. However, since they infer information on the behaviour of other agents only indirectly in hindsight,  fluctuations due to their decisions are unavoidable. They will be in the main focus of this paper.

The minority mechanism that we have just described has some universal features as it underlies a number of optimization problems in dynamical systems whenever it is favorable to belong to the minority.
It is often formulated as a game, the minority game, whose players are the agents. Known as a prototype of the minority game is the El Farol bar problem \cite{Arthur_EL_Farol, coolen_minority_2005}, for which ref.~\cite{challet_shedding_2004} provided  a deeper understanding in terms of statistical physics. Different market mechanisms have been described in terms of minority games in \cite{challet_modeling_2000}. There the spectrum of agents reached from producers to speculators and ``noise traders'', focussing on the information flow between the agents. In \cite{coolen_minority_2005, challet_minority_2000} it is shown that real markets seem to operate near criticality  with marginal efficiency. The stylized description in terms of minority games captures  collective effects, even when the minority game is extended toward models of real markets. Also in relation to financial markets, the minority mechanism takes effect as considered in \cite{challet_criticality_2003}, where the minority game was extended toward a realistic model of the stock market.
After all, models of minority games can be mapped to spin glass models (see e.g. \cite{coolen_minority_2005, challet_modeling_2000}), sharing features of many random variables, quenched disorder and a phase transition between a phase with replica symmetry being broken or restored.

In this paper we focus on the energy market and consider only one type of agents, the arbitrageurs, representing the BRPs of the real market when they get involved in arbitrage. Thus we first translate the dynamics in relation to arbitrage to a minority game. On the formal level, we mainly follow the framework of \cite{challet_modeling_2000}, but extend their versions of minority games to include what corresponds in our applications to a non-vanishing intraday price in combination with risk aversion, moreover external stochastic events in combination with nonlinear payoff functions. As known from complex systems dynamics, stochastic fluctuations in combination with nonlinear functions can lead to unforeseen effects. Furthermore, we use real data for the contributions of BRPs to the exhaustion of reserve energy as a result of arbitrage. Even in this case, remnants of the counterpart to replica symmetry breaking are visible, so this map provides a deeper understanding of the non-monotonic dependence of the fluctuations on the control parameters. In contrast to the perspective from economics, our focus is on the impact of fluctuations around the average arbitrage and their dependence on the model parameters. This dependence is subtle and sometimes counterintuitive.

Before we go into detail in the next sections, we would like to emphasize that arbitrage on the energy market is just one among other possible causes of detrimental effects on the grid stability. As mentioned earlier, large price differences between prices on the intraday market and the reserve power market are favored by strong fluctuations  in renewable energy production, caused by a mismatch of (weather or other) forecast and actual power production. If the production is much lower than expected, it leads to rather high prices with a high risk for retailers. On the other hand, overproduction and oversupply may actually lead to negative energy prices \footnote{What enters our considerations are essentially price differences between intraday and reserve power market prices. We do not explicitly account for negative prices that would be implemented as a penalty for overproduction of renewables.}. This requires a detailed understanding of the time evolution  of spot market prices as a function of deterministic and stochastic effects \cite{keles} (see also the work of \cite{nogales} and \cite{contreras} for time series models, references \cite{tellidou,weidlich,xiong} for agent based modelling and \cite{mariomureddu} for an approach from statistical physics). Thus an alternative path of research is an improved risk management and market design to avoid extreme price occurrences \cite{hagfors}. In view of an increased contribution of renewables to the overall production, fluctuations are inherent and the resulting fluctuations in the prices may be damped, but cannot be completely avoided by appropriate price policies. Therefore also efforts are required on the side of power dispatch, their optimization should be determined under intrinsic uncertainties. Here a possible framework of including uncertainties in the trade on the energy market  is that of aggregative games. These are games of selfish retailers, coupled, however, via common objectives or constraints and communicating directly or indirectly with each other \cite{parise1,parise2}. In contrast, when we focus on the detrimental effect of arbitrage in the framework of the minority game, our retailers are also selfish and share a common motivation to make profit, but have to anti-coordinate to the behavior of the majority to achieve this goal. They do not communicate with each other apart from checking the  success of their own strategies which, of course, does depend on the choice of  other retailers.

The paper is organized as follows.  Section~\ref{sec: minority} presents the general form of minority games with an overview of special cases considered later, followed by a summary of an algorithmic implementation of the minority games. Section~\ref{sec: results}  contains the results. In section~\ref{sec: bounds}, we derive  analytical bounds on the fluctuations suited for a comparison with numerical results. Numerical results are presented first for the minimal version of the minority game to introduce the basic framework (section~\ref{sec: standard}), extended to non-vanishing intraday prices $I$ and nontrivial distributions of the contributions to the imbalance of power (section~\ref{sec: nonvanishing}), the impact of nonlinear price functions in combination with noise (section~\ref{sec: nonlinear}), and the implementation of some degree of risk aversion (section~\ref{sec: GC}). We conclude with a proposal of countermeasures to protect the market from arbitrage (section~\ref{sec: measures}) and summarize some features, which should survive more realistic modelling of the energy market in section~\ref{sec: summary}.

\section{The Energy Market in Terms of Minority Games}\label{sec: minority}
We  distinguish only between two types of energy markets,  the intraday or spot market and the reserve power market. Our agents act on both types. The markets themselves are described by a few parameters and functions.

{\bf Parameters.} The parameters are the total number of BRP-parties $N$, represented by individual agents, who  behave as arbitrageurs, the total amount of power available for arbitrage $W$, where each of our agents $i\in\{1,...,N\}$ has access to with an amount of power $w_i$, and the price $I$ on the intraday market, here kept fixed over time.

{\bf Functional dependencies and distributions.} The reserve power cost function $R$ has to be taken from real data or specified as a (linear or nonlinear) function of the required power. Ongoing fluctuations in the energy balance are caused by differences between forecast and actual consumption or estimated and real production due to fluctuations in renewables. They  are altogether captured by noise $\eta$ of various types and strengths. This ``noise'' adds upon imbalances evoked by our agents, the latter are in the main focus of this paper. A further characteristic is the ``information'' that is in principle available on the market. It may refer to the real history of trade or forecasts on weather, consumption, or production. All this information will be implemented in a highly stylized way as different options of integer numbers $\mu=1,2,..,P$. To each value of $\mu$, a value of either $+1$ or $-1$ can be assigned (representing the decisions $\pm 1$), leading to $2^P$ possible strings of $\pm1$, which later will serve as a ``pool'' of strategies, to be defined below.
Thus information enters only indirectly in the choices of strategies. Here the information represented as $\mu$ is treated as a random variable, drawn uniformly at random from $1, \dots, P$ for any given instant in time.

{\bf Agents, the players of the game.} Our agents are the players of the minority games. They represent BRP-parties, but do not share all activities of general traders (whose aim is to trade such that upcoming or existing imbalances get balanced, while profiting from regular trade of energy). Our agents are arbitrageurs with two possible decisions $a_i=\pm1$. According to our convention, for $a_i=+1$, the agent sells energy on the intraday market, while not feeding enough power into the grid. (This should mimic a situation where the agent tries to buy reserve energy at a price $R<I$ and trades energy on the intraday market as an uncovered sale.) For $R<I$, his payoff per unit of power\footnote{His total payoff is given by $w_i \times u_i$.} $u_i$, given by
\begin{equation}\label{eq1}
u_i:=a_i (I-R),
\end{equation}
will be positive, for $I<R$ and otherwise the same behavior, his payoff will be negative. In contrast, the decision $a_i=-1$ represents a situation, in which the agent buys energy on the intraday market and feeds too much energy into the grid. This is beneficial (detrimental) for him if $I<R$ ($I>R$), leading to $u_i>0$ ($u_i<0$), respectively. Independently of whether the agent has a positive or negative payoff, his decisions can contribute to an imbalance of the market.

\begin{figure}
    \centering
    \subfigure{
     \includegraphics[width = 0.45 \textwidth]{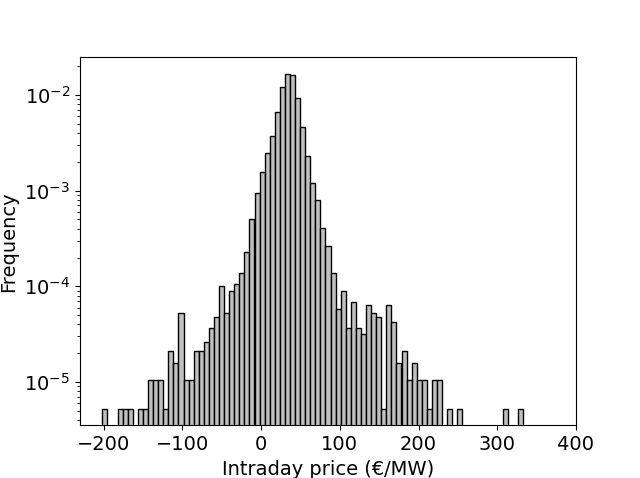}
     \llap{\parbox[b]{17.0cm}{\textbf{\hspace{12.9cm}(a)}\\\rule{0ex}{5.2cm} }}
    }
    \subfigure{
     \includegraphics[width = 0.45 \textwidth]{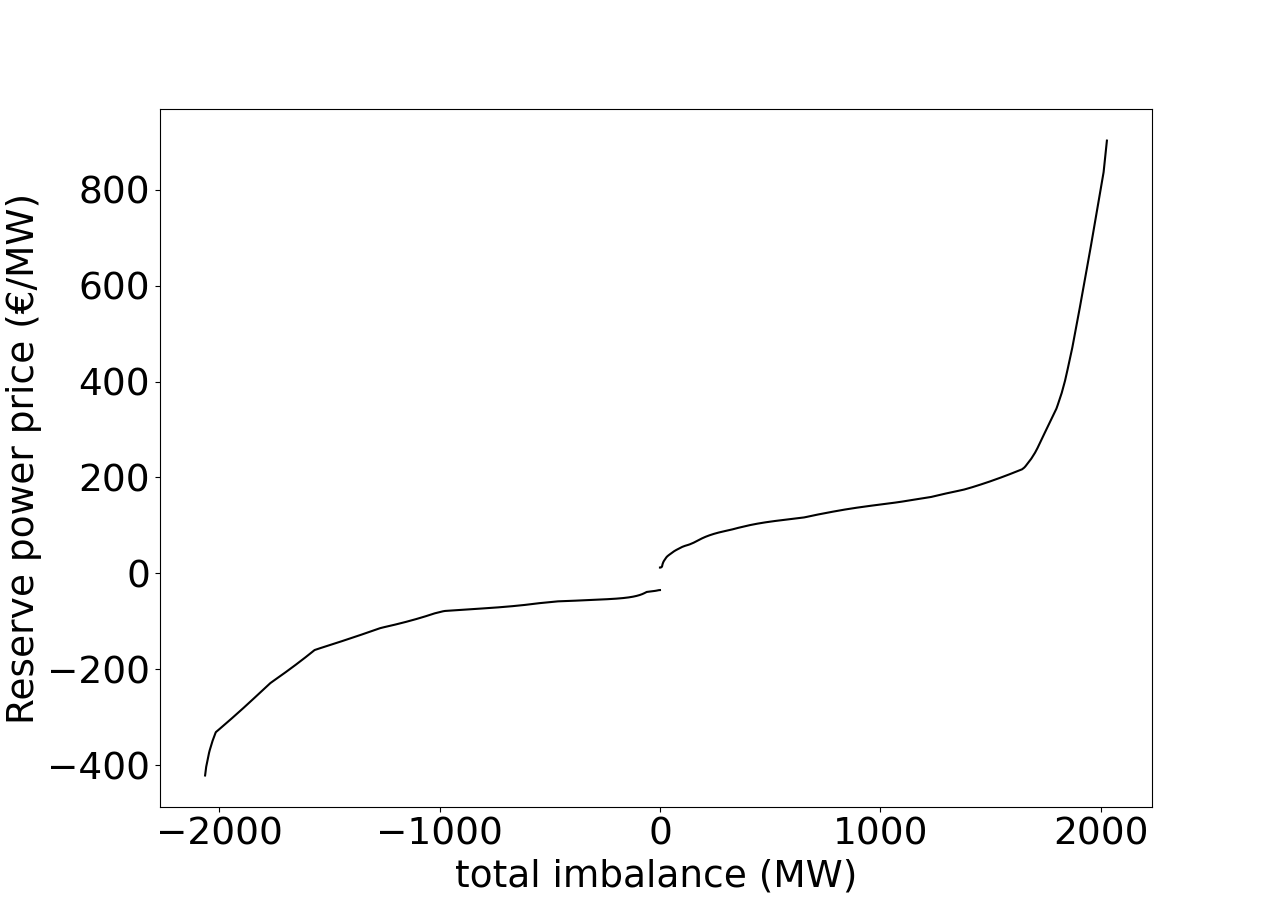}
     \llap{\parbox[b]{17.0cm}{\textbf{\hspace{12.9cm}(b)}\\\rule{0ex}{5.2cm} }}
    }
    \caption{(a) Histogram of the average price on the intraday market over all 15-min intervals (weighted over the total amount of power during these intervals), between 01-01-2017 00:00 and 07-11-2017 24:00, log y-scale.
    Data are chosen from \cite{epex_spot}. (b)   The reserve power price (for secondary reserve) as a function of the size of the total imbalance, on June 1st, 2020, 16:00-20:00 (data are taken from \cite{Regelleistung}). The shape is typical, however, numerical values differ between the four-hour intervals. In particular it can happen that prices are significantly lower than the ones displayed here, leading to dangerous situations.}
    \label{fig: intraday_histogram}
\end{figure}

If we term the total imbalance or shortage of power, caused by the actions of all agents, ``arbitrage'' $A$, the arbitrage is given as
\begin{equation}
A=\sum_{i=1}^N w_i a_i\;.
\end{equation}
The weights $w_i$ with which the agents contribute to the arbitrage need not be chosen uniformly, but from some distribution $P_w(w_i)$. The arbitrage $A$ adds upon other sources of imbalance, summarized under $\eta$. (Note that if $A+\eta$ is positive, some agents can actually reduce the total imbalance by playing $a_i=-1$, and vice versa.)
The price of the reserve power is a function of the total imbalance $A+\eta$ (see section~\ref{sec: introduction}), denoted as $R(A+\eta)$.  Therefore, more precisely, including this in Eq.~\ref{eq1}, the pay-off (per unit of power) for an agent $i$ is given by:
\begin{align} \label{eq: pay-off}
    u_i = a_i \cdot[I - R(A + \eta)].
\end{align}
Since the reserve power is activated according to the merit order, a larger imbalance means that more expensive reserve power is activated. Thus, the reserve power price is increasing when the imbalance grows: $\frac{\mathrm{d} R(x)}{\mathrm{d} x} > 0$ (discussed in more detail in~\ref{app: reserve_deriv}). Figures~\ref{fig: intraday_histogram}(a)  and~\ref{fig: intraday_histogram}(b)  show historical values of $I$ and the typical shape of the reserve power price function $R$, respectively.

{\bf Strategies of the game.} Aspects of game theory enter in how the agents take their decisions. The decisions are based on strategies $s^i(\mu)=a_i\in\{\pm1\}$, $\mu\in\{1,...,P\}$. These are maps from information coded in $\mu$ to decisions $a_i=\pm1$. In this section we give only an overview of the different choices that we consider later in more detail. Each agent can have $S\ge1$ strategies, which need not be homogeneously chosen between the agents, in general they will differ. Strategies could be intrinsically time-dependent. However, we consider those which themselves are constant over time, while the agents may switch between different choices. Moreover, they can be deterministic or stochastic. In section~\ref{sec: bounds} we consider stochastic strategies where each agent $i$ chooses $a_i = 1$ with probability $p_i$ and $a_i = -1$ with probability $1 - p_i$ (independently of $\mu$). If strategies are chosen deterministically, those which are available to the agents are chosen as a certain subset of the pool of all $2^P$ strategies\footnote{The number of possible strategies $2^P$ equals the number of maps from the information coded in $\mu = 1, \dots, P$ to the decisions~$\pm 1$.}. The subset can be chosen uniformly randomly or with a certain bias, the latter case is considered in section~\ref{sec: nonvanishing}. In the course of time, individual agents can switch between the $S$ strategies that had been selected from the pool of $2^P$ ones, depending on their learning process. During the learning phase they evaluate the success of all their strategies in the past, keeping track of their evaluations, recursively defined according to
\begin{equation} \label{eq: update_evaluations}
U^{t+1}_{s^i}= U^t_{s^i}+s^i(\mu^t)\cdot[I-R(A^t + \eta^t)]
\end{equation}
with $s^i(\mu^t)$ denoting the decision of agent $i$ according to strategy $s$, given information $\mu$ at time $t$, so $s^i(\mu^t)\cdot[I-R(A^t + \eta^t)]$ denotes the payoff at time $t$ in the past. In comparing the score of his different strategies, the agent uses for all of them (usually for both if $S=2$) the actual value $A^t$ that was measured when he decided in favor of one strategy at time $t$. In the evaluation he therefore neglects the change in $A^t$ caused by himself if he had chosen another strategy than the actually chosen one. (This error of the order of $1/N$ need not be negligible in certain cases.) At time $t+1$ he chooses the strategy that would have been the best one in hindsight according to his evaluation. Additionally, in section~\ref{sec: GC} we will implement a certain degree of risk aversion by a constraint that an agent only takes a decision if the expected payoff lies above some threshold $\epsilon$, otherwise he refrains from playing at this time step.

{\bf Observables.} Our main observables are the expected amount of arbitrage $\langle A\rangle$ and the variance in the amount of arbitrage $\sigma_A^2 \equiv \langle A^2 \rangle  - \langle A \rangle^2$, averaged over time, as a function of the model parameters $N,P$, risk aversion $\epsilon$, price function $R$, and the choice of the external imbalances $\{\eta^t\}$.
We measure the resulting distribution of $A$, and, importantly, the associated fluctuations $\sigma_A$ and their scaling. The main interest is in the dependence of $\sigma_A$ on the model parameters. Large fluctuations may lead to outliers in the amount of arbitrage and induce an exhaustion of the reserve energy.

{\bf Agent Based Modelling of Minority Games.} Apart from the analytical derivation of bounds on the fluctuations in section~\ref{sec: bounds}, the steps of the agent-based-modelling can be summarized in the following algorithm:
\begin{itemize}
\item Initialization and fixing the choices\\
Choose the parameters: the number of agents $N$, the number of patterns of information $P$, the intraday price $I$, the number of strategies $S$ and the price function $R(x)$. Choose the distribution of external noise $P(\eta)$, the distributions of weights $P_w(w_i)$, the updating rules for the evaluations $U^{t+1}_{s_i}$ (we will here only consider the updating rule given in Eq.~\ref{eq: update_evaluations}) and set the initial values for $U_{s_i}^{t = 0}$ (here we always set them to zero). Choose furthermore the value of the bias in the strategies $p$ and assign (in general different) subsets of $S$ strategies to each agent.
Each strategy $s$ is drawn at random from the space of $2^P$ possible strategies, with the bias $P(s(\mu) = 1) = p$ and $P(s(\mu) = -1) = 1 - p$. Keep these sets of strategies for each agent fixed over the simulation time.
\item Learning phase\\
All agents  simultaneously update their evaluation $U_i$ at time $t+1$, based on the measured value of $A^t=\sum_i w_ia_i^t$ to determine the most promising strategy $s_i^{best}$ in hindsight (the strategy out of their subset of strategies with the highest evaluation).
The signal $\mu^{t+1}$ is drawn uniformly from $1, \dots, P$. All agents use the same signal $\mu^{t+1}$, select the most promising strategy and decide accordingly: $a_i=s_i^{best}(\mu)$ at time $t+1$. Repeat the learning steps until the observables converge. The specific stopping criterion that we use is as follows: Calculate $\sigma_A$ both for the last quarter of time-steps and for the third quarter of time-steps. If the difference between these values is less than $0.1 \%$, stop the simulation and calculate observables over the last half of all  time-steps. Otherwise continue.
\item Measurement of observables\\
Calculate the mean value of arbitrage $\langle A\rangle$ as $\sum_{t=t_0}^{t_0+T} A^t$ from the last $T$ measurements as well as the histogram of $A$ over the time interval $T$. Calculate the corresponding fluctuations $\sigma_A$.
\item Gain statistics for the measurements\\
Repeat the whole procedure so far, including a new initialization of the strategies, for 100 times and average $\langle A\rangle$ and $\sigma_A$ over the hundred iterations. (As we shall see, depending on the ratio of $P/N$ the results may depend on the initialization.)
\end{itemize}

\section{Results} \label{sec: results}
\subsection{Fluctuations of arbitrage: Analytical bounds in limiting cases}\label{sec: bounds}
The pay-offs and possible decisions given in the previous section define a game. To gain some insight into the structure of the pay-offs and its influence on the agents, we consider a version where the game is played only once (in reality the game is repeated every 15 minutes, which we will investigate in more detail in section~\ref{sec: standard}).
To find the expected behaviour of the BRPs, we first try to find Nash equilibria (following a standard game-theoretical procedure \cite{tadelis_game_2013}). These are sets of strategies where no agent has an incentive to deviate from his strategy. In this sense, Nash equilibria are  'stable' solutions of the game.

We consider strategies where each agent $i$ chooses $a_i = 1$ with probability $p_i$ and $a_i = -1$ with probability $1 - p_i$. If $p_i$ equals either $0$ or $1$ the choice is deterministic. Assuming that an individual agent's contribution is small compared to the total imbalance, that is $w_i/A = \mathcal{O}(\frac{1}{N})$ for all $i$ (while keeping $R(A + \eta) = \mathcal{O}(1)$), then  we can neglect the individual agent's contribution to $A$ at large $N$. A set of $\{ p_i \}$ is a Nash equilibrium if no agent has an incentive to change $p_i$. Looking at Eq.~\ref{eq: pay-off}, this is the case when either:
\begin{enumerate}
    \item $I - R(\sum_i w_i + \eta) > 0$ and all $p_i = 1$, \vspace{2mm}
    \item $I - R(- \sum_i w_i + \eta) < 0$ and all $p_i = 0$, \; \text{or}, \vspace{2mm}
    \item $\langle   I - R(A + \eta) \rangle = 0$.
\end{enumerate}
Cases 1 and 2 occur when the difference in intraday price and reserve energy price is so high that there are simply not enough agents to abuse all of the arbitrage opportunities. They correspond to all agents performing arbitrage, respectively by selling or by buying power on the intraday market. The first case corresponds to the aforementioned events in June 2019. Such extreme amounts of arbitrage are rare. Furthermore, measures have been taken to reduce the incentives for the arbitrageurs, reducing the likelihood of case 1 to occur. Case 3 requires less extreme differences in price, and can be expected to occur more often than the other two cases. It is also the more interesting case. Any set $\{ p_i \}$ which leads to this equation being satisfied corresponds to a Nash equilibrium. In general there are many solutions to these equations, each usually corresponding to a different strength of fluctuations around the mean amount of arbitrage and thus to different levels of anti-coordination. Since there are many Nash equilibria, the identification of Nash equilibria as such does not tell the whole story, as it is not clear to which equilibrium interacting agents will converge, or whether they would even converge to an equilibrium at all. Nevertheless it is useful to identify two extremes:
\begin{itemize}
    \item Perfect anti-coordination. \\
Agents choose deterministically, with $p_i$ equal to either $0$ or $1$, such that $I - R(A + \eta)$ is always equal to zero\footnote{Since we assumed each agent's weight $w_i$  to be of order $1/N$, for large $N$ the total amount of arbitrage $A$ is a real number, which can be adjusted to satisfy the equation.}. It requires the agents to separate themselves into two groups, one group playing $a = 1$ and the other group playing $a = -1$. The sizes of the groups must be such that $A$ solves $I = R(A + \eta)$.
    \item No anti-coordination. \\
    An equilibrium which requires no (anti-)coordination between agents can be found assuming the strategies for all agents are homogeneous, that is $p_i = p$ for all $i$. One then needs to find $p$ such that  $\langle   I - R(A + \eta) \rangle = 0$.
\end{itemize}
From the agents' point of view, in either case their pay-off is zero up to corrections of order $1/N$. Keeping $N$ finite and looking at these corrections, they give an expected total pay-off to all agents of $\langle (I - R(A + \eta))\sum_i a_i \rangle  = \langle A(I - R(A + \eta)) \rangle$, which is the correlation between $A$ and $(I - R(A + \eta))$. Since $R(A + \eta)$ is increasing with increasing $A$, this pay-off is only zero if the variance of $A$ is zero, and the pay-off is negative otherwise. The precise value depends on the degree of anti-coordination that the agents achieve. These pay-offs are small (the average pay-off is of order $1/N$), but for finite $N$ one does expect some tendency for the agents to move towards lower variances, and thus achieve some level of anti-coordination. In the 'perfect anti-coordination' equilibrium, the pay-off is exactly zero. From the agents' point of view this is optimal in the sense that arbitrage is maximally exploited just to the limit from which on they make no longer profit. However, it requires internal organization between the agents, and it is not clear how agents would reach this state without explicit agreements. The 'no anti-coordination' solution requires no organization between the agents. In general we expect agents to reach some intermediate position between these extremes. To study the extent to which agents can learn to anti-coordinate, we need to give them some explicit learning dynamics. We will introduce this in section~\ref{sec: standard}.\\

{\bf Estimates of the order of magnitude of fluctuations.}
We can find an order of magnitude of the strength of the fluctuations by looking at the variance predicted by the 'no anti-coordination' Nash equilibrium. Let us denote the amount of arbitrage that removes all incentives for further arbitrage by $A^\ast$. That is, $A^\ast$ is such that the price of reserve power equals the price of power on the intraday market, $R(A^\ast + \eta) = I$. If we approximate $R$ as linear around $R(A^\ast + \eta) = I$, that is $R(A + \eta) \sim I + c(A - A^\ast)$ for some $c$, then a simple solution can be found. If fluctuations around $A^\ast$ are small, this is a good approximation. If the reserve price function is linear, one simply has to set the mean amount of arbitrage $\mu_A \equiv \langle A \rangle = A^\ast$, giving  $\langle a \rangle = 2p - 1 = A^\ast/\sum_i w_i$ and variance:
\begin{align} \label{eq: order_of_mag}
    \sigma_A^2 \equiv \langle A^2 \rangle - \mu_A^2 = \frac{W^2}{N/X} \times (1 - \mu_A^2/W^2) = \mathcal{O}(N) \,.
\end{align}
Here we defined the total power available for arbitrage by
\begin{equation}
W \equiv \sum_i w_i,
\end{equation}
(which is $\mathcal{O}(N)$) and
\begin{equation}
X \equiv \big(\frac{1}{N} \sum_i w_i^2 \big) /\big(\frac{1}{N} \sum_i w_i \big) ^2\;,
\end{equation}
which measures the non-uniformity of the distribution of weights $w_i$. (Here expectation values $\langle \dots \rangle$  refer to the distribution of $a_i$.) For a uniform distribution of weights $X = 1$, while for a non-uniform distribution it is always greater than $1$ (for example, for exponential distributions $X = 2$). $N/X$ is an effective number of agents contributing to the fluctuations: if $N'$ agents have (uniform) non-zero weight (and the rest has weight zero), then $N/X = N'$. We note that $W = \mathcal{O}(N)$,  giving also $\sigma_A^2 = \mathcal{O}(N)$. In fact, adding an additional arbitrageur always increases $\sigma_A$, disproportionally so for arbitrageurs with relatively high weight.\footnote{To be precise, adding an arbitrageur with weight $w_j \ll W_{tot}$ and expanding the variance $\sigma_A^2$ to first order in $w_j/W_{tot}$, we find that it is multiplied by a factor $1 + w_j \frac{2\mu_A^2}{W(W^2 - \mu_A^2)} + \frac{w_j^2}{\sum_i w_i^2} > 1$.}\\
To estimate the significance of these fluctuations to arbitrage on the energy market, we need some estimate of $A^\ast$ and of the weights $w_i$ (note that the variance is actually independent of the coefficient $c$). In \ref{app: OOM} we use the description of the events of June 2019 given by  \cite{50_hertz_untersuchung_2019} to make such an estimate, which leads to an order of magnitude $\sigma_A/\mu_A \approx 0.4-0.5$ if we insert the data into Eq.~\ref{eq: order_of_mag}. The fact that these fluctuations are not much smaller than the expected amount of arbitrage $\mu_A$, means that they contribute significantly to the risk of exhausting the reserve power. What is more, in section~\ref{sec: nonlinear} we will find that for a nonlinear price function, the expected amount of arbitrage itself also changes depending on the magnitude of the fluctuations. These findings make clear that the fluctuations require a careful study.

In summary, the estimate according to Eq.~\ref{eq: order_of_mag}, derived without any learning process, will be used for comparison with situations in which arbitrageurs have reduced the fluctuations due to learning.

\subsection{The standard minority game for the energy market}\label{sec: standard}
\begin{figure}
    \centering
    \subfigure{
    \includegraphics[width = 0.45 \textwidth]{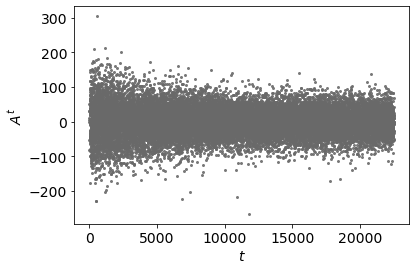}
     \llap{\parbox[b]{17.0cm}{\textbf{\hspace{13.55cm}(a)}\\\rule{0ex}{5.2cm} }}
    }
    \subfigure{
     \includegraphics[width = 0.45 \textwidth]{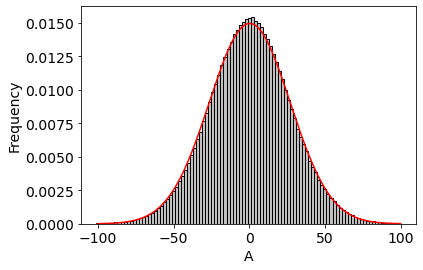}
     \llap{\parbox[b]{17.0cm}{\textbf{\hspace{13.55cm}(b)}\\\rule{0ex}{5.2cm} }}
    }
 \caption{ (a) The evolution of $A^t$, for $N = 4100$, $P = 2050$ and $S = 2$, showing its decrease as a function of time. (b) A histogram of $A^t$ (for the same parameters) over $10^6$ time-steps. The solid line shows a Gaussian distribution with the same variance as the histogram.}
    \label{fig: scatter_A}
\end{figure}

Let us first reproduce known results and discuss the 'standard' minority game, which uses a special case of the pay-off described in section~\ref{sec: minority}. This will lead to an understanding of the basic structure of minority games resulting from the learning dynamics.  We consider the following case (studied in \cite{challet_modeling_2000}):\\
$\bullet$ The weights $w_i$ are uniformly equal to $1$, $\bullet$ the intraday price $I = 0$,   $\bullet$ the imbalance due to other causes $\eta = 0$, $\bullet$ the reserve price is given by $R(A + \eta) = R(A) = A/N$.
This gives the payoff:
\begin{align}
u_i = - a_i A/N \,.
\end{align}
We note that setting $R(A) = cA/N$ for some constant $c$ simply multiplies the payoffs by $c$ and does not change the behaviour of the agents compared to the $c = 1$ case.
We furthermore note that the authors of~\cite{coolen_minority_2005} study the payoff $u_i = - a_i \, \text{sgn}(A)$ instead.
We follow the learning dynamics introduced in section~\ref{sec: minority}. The strategies are drawn with zero bias: In any given strategy, $a = 1$ has the same probability of occurrence as $a = -1$. \\
Let us now see the type of behaviour resulting from these dynamics. As a starting point we consider particular values of the parameters: $N = 4100$, $P = 2050$ and $S = 2$. The time evolution of $A^t$ is shown in Fig.~\ref{fig: scatter_A}(a). The precise value of $A^t$ at a given time-step is highly unpredictable. Plotting a histogram of all the values attained by $A^t$, Fig.~\ref{fig: scatter_A}(b) shows that the values of $A^t$ follow a Gaussian distribution. A general trend in the time evolution of $A$ (Fig.~\ref{fig: scatter_A}(a)) can furthermore be observed: For small $t$, values of $A$ that are far from the average value are relatively common; as time evolves the agents learn to anti-coordinate, and such values become more rare.
\begin{figure}[h]
    \centering
    \includegraphics[width = 0.45 \textwidth]{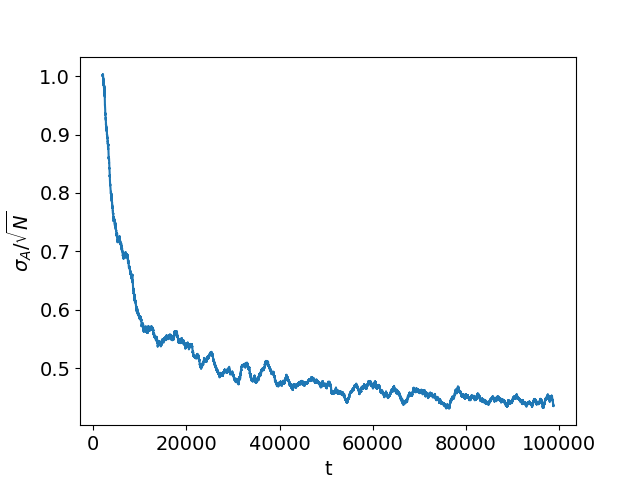}
    \caption{Evolution of $\sigma_A$ for $N =  4100$, $P = 2050$ and $S = 2$, showing that the agents learn to anti-coordinate; $\sigma_A$ is calculated by using a running average over $2000$ time steps, only initially it is equal to $\sqrt{N}$ as naively expected. }
    \label{fig: sigma_evolution}
\end{figure}
We can quantify this effect by interpreting the histogram shown in Fig.~\ref{fig: scatter_A}(b) as a probability distribution from which $A^t$ are drawn independently. Indeed, in ref.~\cite{challet_minority_2000} it is shown that for a wide range of parameters, including the ones chosen here, $A^t$ are effectively behaving like random variables.
\footnote{This is despite the fact that once the initial conditions have been chosen (i.e. the strategies have been drawn), the process is completely deterministic. The randomness is \textit{effective}, caused by the large amount of agents interacting disorderedly with each other. Such an effect occurs more often in disordered systems \cite{castellani_spin-glass_2005}.}. Averages over the distribution $P(A)$ from which $A^t$ are drawn can then be calculated by performing time averages:
\begin{align}
    \langle f(A) \rangle_{P}  \equiv \frac{1}{T}\sum_{t = t_0}^{t_0 + T} f(A^t) \,,
\end{align}
for large $T$ and $t_0$. Time instants $t_0$ and $t_0 + T$ determine the time interval over which the average is taken, and the distribution $P(A)$ is assumed to remain (approximately) the same throughout this interval\footnote{This is true for $t_0 \gg T$, as  shown in \cite{challet_modeling_2000} that the system eventually approaches a fixed distribution $P(A)$.}. The statement that values of $A^t$ become closer to the mean as the agents learn to anti-coordinate,  translates then to the statement that the variance $\sigma_A^2 \equiv \langle A^2 \rangle - \langle A \rangle^2$ decreases as time progresses.

The time evolution of $\sigma_A$ is shown in Fig.~\ref{fig: sigma_evolution}. At the start of the game, agents have not yet learned to anti-coordinate, and they behave according to the 'no anti-coordination' Nash equilibrium (which would give $\sigma_A/\sqrt{N} = 1$). As time goes on, they learn from the results in the past to predict $A$ to some extent: The agents effectively achieve some degree of anti-coordination, and thereby decrease the fluctuations in $A$.

\begin{figure}
    \centering
    \subfigure{
    \includegraphics[width = 0.45 \textwidth]{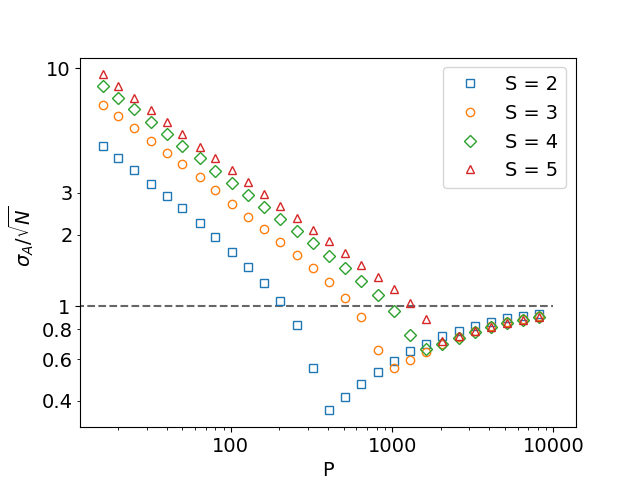}
    \llap{\parbox[b]{17.0cm}{\textbf{\hspace{13.00cm}(a)}\\\rule{0ex}{5.2cm} }}
   }
   \subfigure{
   \includegraphics[width = 0.45 \textwidth]{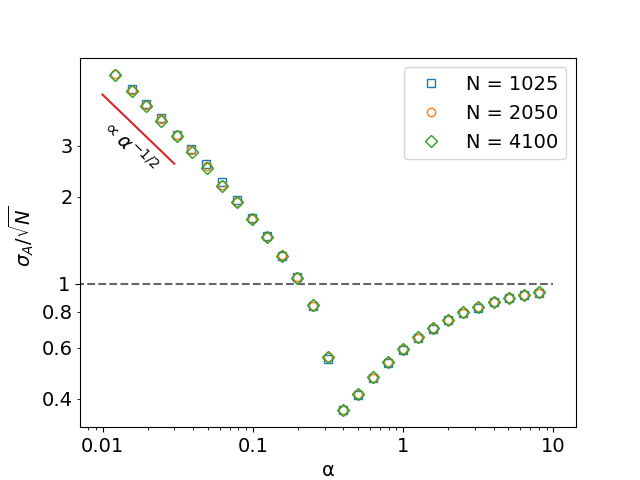}
    \llap{\parbox[b]{17.0cm}{\textbf{\hspace{13.00cm}(b)}\\\rule{0ex}{5.2cm} }}
   }
    \caption{(a) Scaled fluctuations of arbitrage for $N = 1025$, as a function of  $P$ for different $S$. Each data-point is an average over $100$ samples. The dashed line corresponds to the hypothetical case, where agents would not have learned at all, given by the 'no anti-coordination' Nash equilibrium. (b) Scaled fluctuations for $S = 2$. The curves for different $N$ collapse on each other if they are shown as a function of $\alpha \equiv P/N$.}
    \label{fig: different_S}
\end{figure}

The extent to which they anti-coordinate depends on the values of the parameters. Even for this simplest case, we have three different parameters to tune: $N, P$ and $S$. To investigate the dependence of the collective behaviour on these parameters, we run the minority game for different values of $N$, $P$ and $S$. For any given set of parameters, we run the minority game until $\sigma_A$ converges to a constant value (see section~\ref{sec: minority}). The values of $\sigma_A$ to which the system converges are shown in Fig.~\ref{fig: different_S}(a) for fixed $N$, and varying values of $P$ and $S$. For each of the values of $S$, the standard deviation $\sigma_A$ to which the system converges shows a non-trivial dependence on $P$. For high $P$, the agents fail to reach any anti-coordination at all, and behave equivalently to the 'no anti-coordination' Nash equilibrium. As $P$ is lowered, the anti-coordination that the agents achieve increases, until finally a turning point is reached, where $\sigma_A$ shoots up, eventually reaching values much higher than they would have achieved if they had not learned at all.

\begin{figure}
    \centering
    \subfigure{
   \includegraphics[width = 0.45 \textwidth]{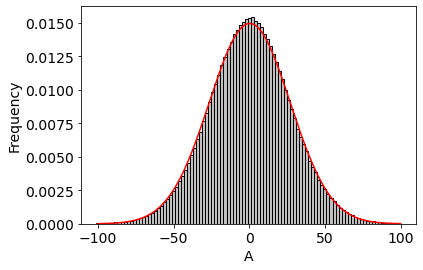}
      \llap{\parbox[b]{17.0cm}{\textbf{\hspace{13.55cm}(a)}\\\rule{0ex}{5.2cm} }}
    }
    \subfigure{
   \includegraphics[width = 0.45 \textwidth]{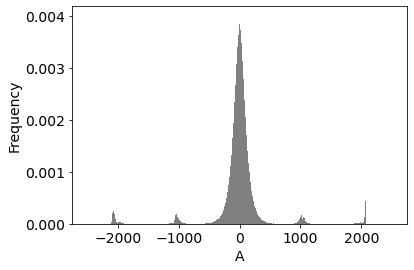}
     \llap{\parbox[b]{17.0cm}{\textbf{\hspace{13.45cm}(b)}\\\rule{0ex}{5.2cm} }}
    }
    \caption{Histograms of $A^t$ (after convergence) for (a) $\alpha = 1 > \alpha_c$ and (b) $\alpha \approx 0.0076 < \alpha_c$; $N = 4100$, averages over $10^6$ time-steps.  For $\alpha < \alpha_c$ (b), the distribution is strongly non-Gaussian, in contrast to the $\alpha > \alpha_c$ case (a).}
    \label{fig: histogram}
\end{figure}

Different values of $N$ are easily incorporated, as $\sigma_A/\sqrt{N}$ depends only on the ratio $\alpha \equiv P/N$ \cite{ coolen_minority_2005, challet_modeling_2000}; this is shown for $S = 2$ in Fig.~\ref{fig: different_S}(b). The same holds for $S > 2$.
The non-monotonic behaviour is known to be due to a phase transition at $\alpha_c \approx 0.34$ (\cite{ coolen_minority_2005,challet_modeling_2000}). The phase transition corresponds to a type of transition known from the theory of disordered systems, called replica-symmetry breaking \cite{coolen_minority_2005,castellani_spin-glass_2005}. For $\alpha > \alpha_c$, the distribution from which $A$ is effectively drawn is Gaussian, as in Fig.~\ref{fig: scatter_A}(b) (and repeated in Fig.~\ref{fig: histogram}(a)). The low-$\alpha$ phase shows qualitatively different behaviour: The histogram shown in~\ref{fig: histogram}(b) shows that the distribution of $A$ is no longer Gaussian, and has rather extreme outliers. In the context of the reserve power these outliers represent dangerous situations, where the amount of arbitrage is much higher than would be naively expected.

The low-$\alpha$ phase has the further detrimental property that $\sigma_A/\sqrt{N}$ increases with decreasing $\alpha$; Fig.~\ref{fig: different_S}(b) shows that for small $\alpha$, the standard deviation scales as $\sigma_A/\sqrt{N} \sim 1/\sqrt{\alpha}$. Since $\alpha = P/N$, this means that for fixed $P$ the deviations scale as $\sigma_A \sim N$ instead of the expected $\sigma_A \sim \sqrt{N}$. The strength of the fluctuations therefore depends non-trivially on the number of agents $N$. Starting at high $\alpha$ (low $N$), $\sigma_A/\sqrt{N}$ is more or less constant and equal to $1$. Therefore, $\sigma_A$ increases as $\sqrt{N}$, as one would initially expect. Increasing $N$ further decreases $\alpha$ and actually makes $\sigma_A$ growing slightly slower ($\sigma_A/\sqrt{N}$ decreases, as seen in Fig.~\ref{fig: different_S}(b), due to anti-coordination by the agents). Increasing $N$ even more, the phase transition is reached, until eventually $\sigma_A \sim N$.

\subsection{Non-vanishing intraday prices and non-uniform participation of arbitrageurs} \label{sec: nonvanishing}
To apply the ideas developed for the minority game to reserve power arbitrage, we need to generalize the game to more realistic assumptions.
First of all, let us consider some realistic parameter values. Formally we follow in this section ref.~\cite{challet_shedding_2004}, where, however, the meaning of $I$ is quite different. In ref.~\cite{challet_shedding_2004}, the El Farol bar problem is studied (the prototype realization of a minority game), where $W = N$ players consider going to the El Farol bar, and $I$ plays the role of the maximum number of visitors of the El Farol bar, for which the available space in the bar is still convenient for the visitors. We leave $W$ and $I$  arbitrary, and first take the weights uniformly: $w_i = W/N$. We note that the value of  $W$ simply rescales the amount of arbitrage $A$, and for linear cost-functions therefore does not influence the dynamics. The case discussed in the previous section corresponds to $W = N$ and $I = 0$.

We change the bias of the initial drawing of strategies (as discussed in section~\ref{sec: minority}) such that $a = \pm 1$ occur following the probabilities given by the no anti-coordination Nash equilibrium (section~\ref{sec: bounds}). In this way, if agents have only one strategy ($S = 1$), implying they have no ability to learn, they behave according to the no anti-coordination Nash equilibrium. Here we will choose also $S=2$, for which it is not automatically guaranteed that they  approach this Nash equilibrium. We note that for realistic learning dynamics also the bias itself should be obtained by some learning process; to our knowledge, no such dynamics has been studied so far, although it would amount to an interesting extension. We stick with the procedure given in \cite{challet_shedding_2004}, and change the bias as discussed above, according to the no anti-coordination Nash equilibrium. This corresponds to choosing $a_i = \pm 1$ with probabilities $p=\frac{1}{2} \pm \frac{1}{2} \frac{A^\ast}{W}$. As a reminder, $A^\ast$ solves $I - R(A^\ast + \eta) = 0$. For linear reserve price function $R$, it is equal to the mean amount of arbitrage. The no anti-coordination Nash-equilibrium would give (assuming uniform weights) the standard deviation:
\begin{align} \label{eq: Nash_std}
    \sigma_A = \frac{\sqrt{W^2 - {A^\ast}^2}}{\sqrt{N}} = \frac{W}{\sqrt{N}} \times \sqrt{1 - (A^\ast/W)^2} \,.
\end{align}
For $S = 1$ the agents do not learn, and (due to the bias) behave according to this Nash equilibrium, defined in terms of the probability $p$ in the stochastic strategy which leads to $I - R(A^\ast + \eta) = 0$. For $S \geq 2$ the agents are able to learn, and the fluctuations can deviate from Eq.~\ref{eq: Nash_std}.
The authors of \cite{challet_shedding_2004} show that in the high-$\alpha$ phase the standard deviation still scales exactly with $\sqrt{W^2 - {A^\ast}^2}$, simply replacing the first equality sign in Eq.~\ref{eq: Nash_std} with proportionality:
\begin{align} \label{eq: scaling}
    \sigma_A \propto \frac{\sqrt{W^2 - {A^\ast}^2}}{\sqrt{N}} \,,
\end{align}
where the  proportionality factor depends on $\alpha$. However, while the proportionality is exact in the high-$\alpha$ phase, there is a small correction in the low-$\alpha$ phase.

\begin{figure}
    \centering
    \hspace{-0.45cm}
    \subfigure{
   \includegraphics[width = 0.45 \textwidth]{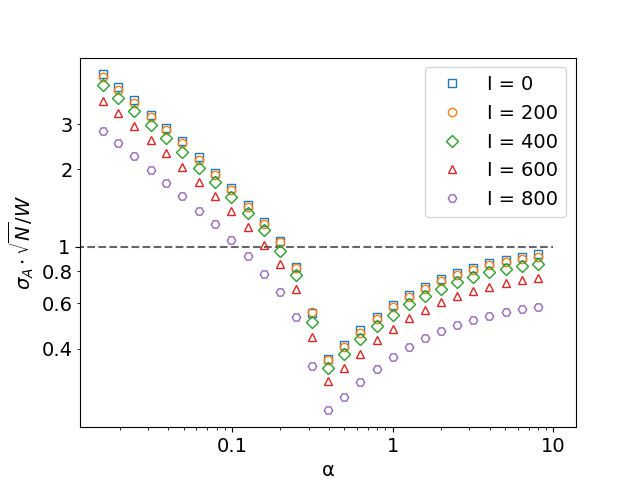}
     \l \llap{\parbox[b]{17.0cm}{\textbf{\hspace{13.0cm}(a)}\\\rule{0ex}{5.2cm} }}
    }
    \subfigure{
   \includegraphics[width = 0.45 \textwidth]{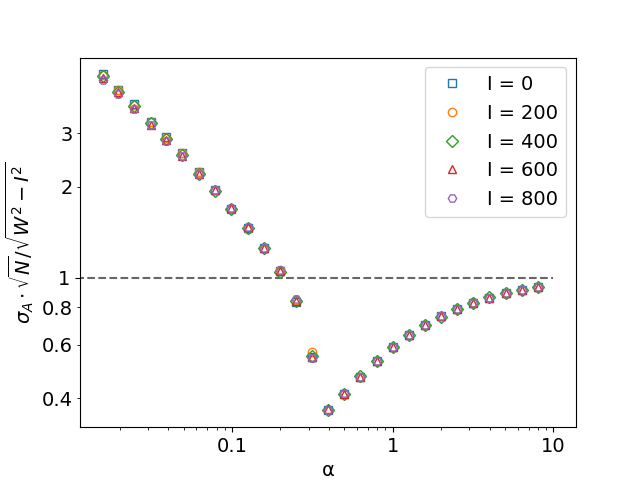}
      \llap{\parbox[b]{17.0cm}{\textbf{\hspace{13.0cm}(b)}\\\rule{0ex}{5.2cm} }}
    }
    \caption{Rescaling of the fluctuations as a function of $\alpha$ for various intraday prices $I$ with $N = 1025$ and $S = 2$, $P=\alpha N$. Each data point is an average over 100 samples. For further explanations see the text.}
    \label{fig: varying_I}
\end{figure}

To illustrate the scaling according to Eq.~(\ref{eq: scaling}), we will use the reserve price function $R(A + \eta) = A$. Solving $I - R(A^\ast + \eta) = 0$ then simply gives $A^\ast = I$. In Fig.~\ref{fig: varying_I}(a) we plot the resulting standard deviation as a function of $\alpha$, for different values of $I$.  The results are nearly  indistinguishable from simply multiplying the standard deviation by $\sqrt{W^2 - I^2}$ (Fig. \ref{fig: varying_I}(b)).

\begin{figure}[h]
    \centering
    \subfigure{
   \includegraphics[width = 0.45 \textwidth]{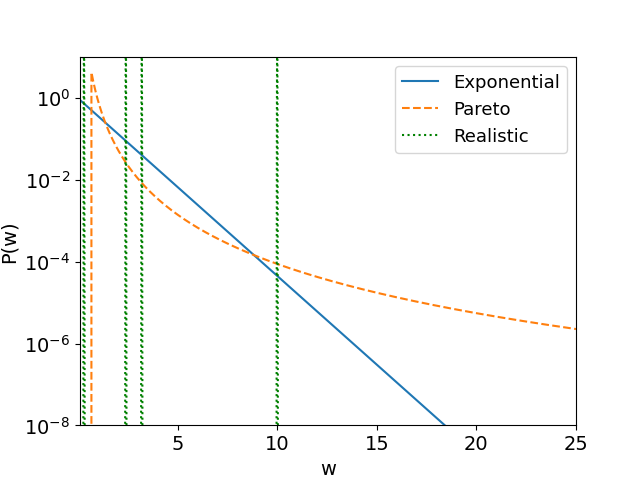}
      \llap{\parbox[b]{17.0cm}{\textbf{\vspace{-0.2cm}\hspace{13.00cm}(a)}\\\rule{0ex}{5.2cm} }}
    }
    \subfigure{
   \includegraphics[width = 0.45 \textwidth]{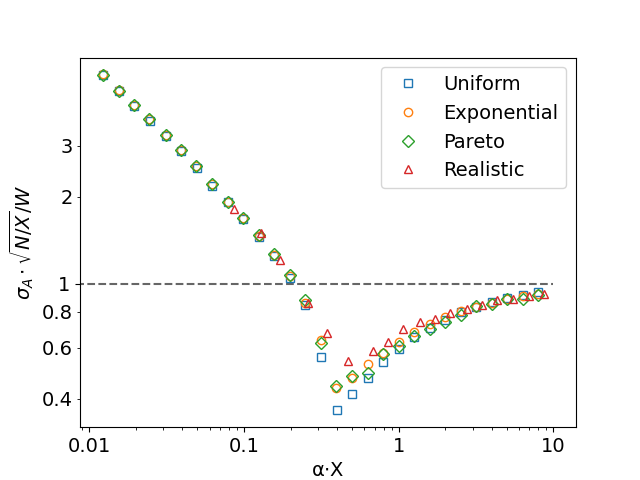}
     \llap{\parbox[b]{17.0cm}{\textbf{\vspace{-0.2cm}\hspace{13.00cm}(b)}\\\rule{0ex}{5.2cm} }}
    }
    \caption{(a) Visualization (with logarithmic y-scale) of different weight distributions: an exponential distribution, a Pareto distribution with $P_w(w) \propto w^{-4}$ and the realistic distribution described in \ref{app: OOM}, all scaled such that $\langle w \rangle = 1$. (b) Rescaled fluctuations for different weight distributions, where the rescaling is prescribed by the 'no anti-coordination' Nash equilibrium estimate. For the realistic distribution, the weights are set exactly according to \ref{app: OOM}; for the other distributions, the weights are drawn independently according to the respective distribution ($N = 4100$). Each data point is an average over 100 samples.}
    \label{fig: weights_00}
\end{figure}

The case with non-uniform weights has been worked out in \cite{challet_minority_2000}. It was found that for weights $w_i$, drawn from a distribution $P_w(w)$, the results are not simply changed through the scaling $\sigma_A \propto \sqrt{X} \equiv\sqrt{\overline{w^2}/\overline{\mbox{$w$\raisebox{0.2cm}{}}}^2}$, (where the overbar denotes an average over $P_w(w)$), which is what the 'no anti-coordination' Nash equilibrium would imply (section~\ref{sec: bounds}). In Fig.~\ref{fig: weights_00}(b) we show the results for an exponential distribution ($X = 2$), the Pareto distribution $P_w(w) \propto w^{-4}$ for $w > 2/3$ (which has $X = 4/3$), and for the realistic distribution described in \ref{app: OOM} ($X \approx 5$). The distributions are furthermore visualized in Fig. \ref{fig: weights_00}(a). Fig. \ref{fig: weights_00}(b) shows that even though the scaling $\sigma_A \propto \sqrt{X}$ is not exact, it still gives a good approximation. This is despite the fact that the realistic distribution has a very small effective number of agents ($N/X \approx 23$). It is remarkable that even in this case the expected finite size effects due to small $N/X$ and the realistic weights of contributions  to the overall power $W$ do not interfere with the underlying structure of  a phase transition.

The scaling of the variance $\sigma_A^2 \propto (W^2 - I^2)/(N/X)$ according to Eq.~\ref{eq: order_of_mag} is not exact. However, in this section we have found  that despite a few quantitative differences, the scaling gives a very good  approximation over a large range of different values. Apparently the choice of bias in the selected strategies (here also for $S\ge2$) is responsible for the success of the  scaling relation. It thus provides a very useful way of understanding the behaviour of the variance of $A$ for a very wide variety of intraday prices $I$ and distributions of weights $\{ w_i \}$ and should  be exploited for optimizing the choice of parameters in realistic market designs.

\subsection{Noise in combination with nonlinear power price functions}\label{sec: nonlinear}

\begin{figure}
    \centering
    \includegraphics[width = 0.5 \textwidth]{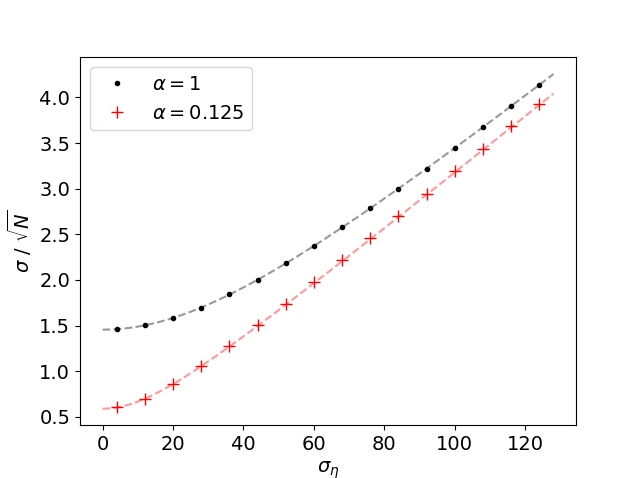}
    \caption{The standard deviation of the total imbalance $A + \eta$, denoted by $\sigma$, for a linear pay-off function. The dashed lines correspond to the prediction that the variance in $A$ and the variance of the noise add linearly, that is $\sigma^2 = \sigma_A^2 + \sigma_\eta^2$, that is, for given $\sigma_\eta$, measured $\sigma_A$, calculate $\sigma$, and compare it to the numerically measured $\sigma$ (crosses and dots).  Gaussian noise $\eta$ with zero mean and varying variance $\sigma_\eta$ has been added with fixed $N = 1025$ and $S = 2$. Each data point is an average over 100 samples.}
    \label{fig: Noise}
\end{figure}

In general, arbitrage is not the only contribution to the total imbalance. Even in the absence of arbitrage, there are BRPs that do not have their portfolio balanced due to deviations of renewable power or consumption from their predictions. As in section~\ref{sec: minority} we denote the imbalance caused by these fluctuations by $\eta$. Since by definition the deviations cannot be predicted by the BRPs, we treat them as a random variable, with zero mean. They are independently drawn for each time-step, from a distribution which has to be specified. They thus introduce noise into the minority game. We will now inspect how this noise affects the dynamics of the agents.

\subsubsection{Linear reserve power price}
First, let us understand how the noise changes the pay-off for the agents. Up to order $1/N$, we can split the expected pay-off of correlated $a$ and $I - R(A + \eta)$ into the product  $\langle a \rangle\langle I - R(A + \eta) \rangle$. The expectation $\langle I - R(A + \eta) \rangle$ thus determines which of the choices $a_i = \pm 1$ gives a higher pay-off. If we assume linear reserve energy price, $R(A + \eta) = I + c(A + \eta - A^\ast)$, the expected pay-off is given by $-\langle c \cdot (A + \eta - A^\ast) \rangle = - c \cdot \big( \langle A  - A^\ast \rangle + \langle \eta \rangle)$. Any noise with mean zero will on average not change the preference between $a_i \pm 1$. If the learning dynamics happens on time-scales longer than that needed for $\langle \eta \rangle$ to average out to zero, one thus expects noise in  $\eta$ not to influence the choices of the agents at all, and to be uncorrelated with $A$. The variance of the total imbalance $A + \eta$ (denoted by $\sigma^2$) would then simply be the sum of the separate variances: $\sigma^2 = \sigma_A^2 + \sigma_\eta^2$. The results of Fig.~\ref{fig: Noise} show that this is indeed the case.

\subsubsection{Nonlinear reserve power price}
It is well known from the physics of complex systems that noise in combination with nonlinear dynamics may have counterintuitive or unforeseen effects, in particular constructive ones as we shall see in this section. Thus, for a nonlinear reserve power price or pay-off function, the role of noise will be analyzed. For a linear price function we had the simplification that $\langle I - R(A + \eta) \rangle = c \langle A + \eta - A^\ast \rangle$ only depends on the mean of $P(A + \eta)$, and thus setting it to zero uniquely determines the mean amount of arbitrage. For a nonlinear price function, $\langle  R(A + \eta) \rangle = \int_{- \infty}^{\infty} \dd x\,R(x) P(A + \eta = x) $ depends on the whole distribution of $A + \eta$. To understand how this changes the outcomes of the game, let us first consider the price function up to quadratic order around $A^\ast$:
\begin{align}
    R(A + \eta) &= I + c_1(A + \eta - A^\ast) + c_2(A + \eta - A^\ast)^2  \\
    \rightarrow \langle R(A + \eta) \rangle &=  I + c_1 \langle A  - A^\ast \rangle + c_2 (\sigma_A^2 + \sigma_\eta^2) \,.
\end{align}
Thus, setting $\langle I - R(A + \eta) \rangle = 0$ shows that, if we consider Nash equilibria of the non-repeated game, the variance in $A$ and $\eta$ shifts the mean of $A$ away from the value that it would have for linear cost-function (which would be $A^\ast)$. There is thus an interaction between the mean and variance of the imbalance $A + \eta$. If $c_2$ is positive the mean of $A$ becomes lower, and vice-versa. Essentially this effect is due to the fact that in the presence of a positive second derivative of the price function, noise lowers the profits of the arbitrageurs, causing them to leave some arbitrage opportunities unexploited (which would have been worth exploiting in the absence of noise).

A look at the example price function of Fig.~\ref{fig: intraday_histogram}(b) shows that for positive values of the imbalance, it is in general the case that the second derivative is indeed positive throughout almost the entire graph \footnote{The actual function changes every four hours. Although it is often similar to the one in Fig.~\ref{fig: intraday_histogram}(b), it occasionally occurs that the second derivative is negative throughout large parts of the graph.}. As shown in \ref{app: reserve_deriv}, if the marginal price of reserve power increases fast enough, the second derivative of the price function is always positive. A positive second derivative of the price function means that noise $\eta$ and variance in $A$ lower the expected amount of arbitrage.

\begin{figure}
    \centering
    \subfigure{
   \includegraphics[width = 0.45 \textwidth]{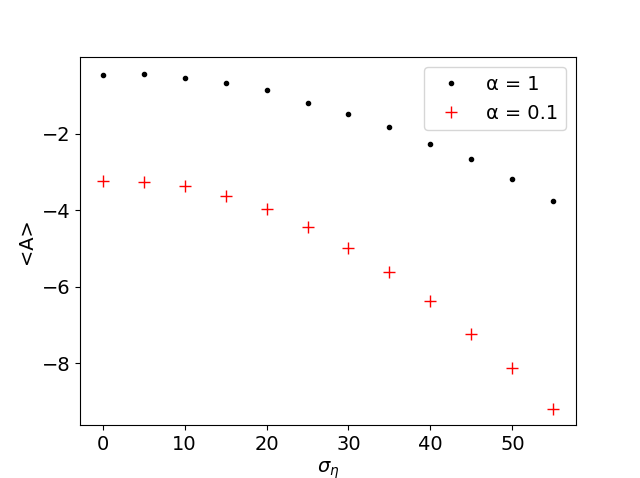}
     \llap{\parbox[b]{17.0cm}{\textbf{\hspace{13.00cm}(a)}\\\rule{0ex}{5.2cm} }}
    }
    \subfigure{
   \includegraphics[width = 0.45 \textwidth]{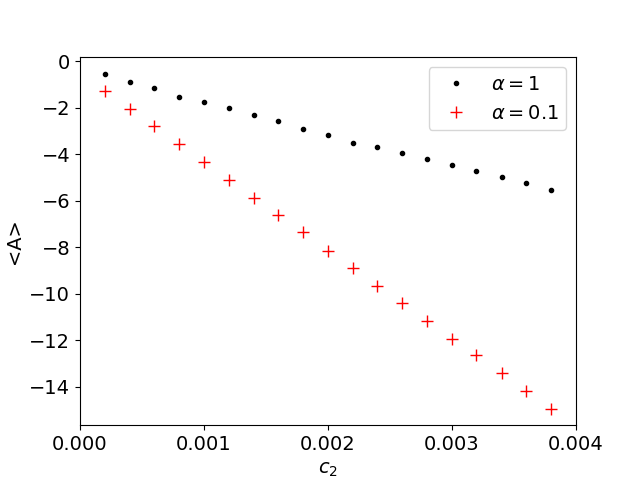}
     \llap{\parbox[b]{17.0cm}{\textbf{\hspace{13.00cm}(b)}\\\rule{0ex}{5.2cm} }}
    }
    \caption{ Decrease of the expected arbitrage $\langle A \rangle$ (a) as the external fluctuations $\eta$ increase,  (b) as the strength of the nonlinearity introduced by $c_2$ increases. Here for the minority game with quadratic price function from Eq.~\ref{eq: quadratic}, for $N = 1025$, $S = 2$.}
    \label{fig: Noise_2}
\end{figure}

To further confirm that a positive second derivative of the price function implies that noise lowers the expected amount of arbitrage, we run the 'basic' minority game from section~\ref{sec: minority} with $N = 500$ and $S = 2$.  This time we add a small quadratic component to the price function:\footnote{We keep the initial drawing of the strategies unbiased, i.e. $s(\mu) = \pm 1$ occur with equal probability when they are initially selected.}
\begin{align} \label{eq: quadratic}
R(x) = x + c_2 x^2 \,.
\end{align}
Fig.~\ref{fig: Noise_2} shows the expected amount of arbitrage $\langle A \rangle$ as a function of the noise (Fig. \ref{fig: Noise_2}(a), for $c_2 = 1/500$) and of $c_2$ (Fig.~\ref{fig: Noise_2}(b), for $\sigma_\eta = 50$) for $\alpha = 0.1$ and $\alpha = 1$. In section~\ref{sec: standard} we found that, even for zero noise, the minority game for $\alpha = 0.1$ gives a large variance of $A$. Consequently, Fig.~\ref{fig: Noise_2}(a) shows a mean amount of arbitrage that is lower than that for $\alpha = 1$. In both cases the total variance of $A + \eta$ increases when the noise strength increases, which in turn means that the mean is decreasing when the strength of the noise increases (Fig. \ref{fig: Noise_2}(a)). (The larger the noise, the lower the mean to satisfy the Nash equilibrium condition $\langle I-R\rangle =0$.) Likewise, Fig. \ref{fig: Noise_2}(b) shows that the expected amount of arbitrage can also be decreased by increasing the non-linearity of the reserve power price function (as determined by the value $c_2$).

Note that the discussion so far holds for arbitrary distributions of noise, as for a quadratic price function only the variance of the noise enters the average pay-off. For general price functions, the entire distributions of the noise and of $A$ are important. However, similar behaviour may hold: In \ref{app: non_linear_cost} we consider a distribution of $A + \eta$ which is symmetric about its mean. We show that on any interval where the second derivative of the price function is positive, a broadening of the distribution (,that is, a shift of the probability mass away from the mean amount of imbalance, e.g. due to noise $\eta$, or due to the distribution of $A$ itself,) increases $\langle R(A + \eta) \rangle$ and therefore decreases the mean amount of arbitrage for which a Nash equilibrium is achieved. The reverse also holds: A negative second derivative increases the mean amount of arbitrage.

\subsection{The impact of risk aversion}\label{sec: GC}

In section~\ref{sec: bounds} we found that the less agents are involved in the game, the lower the severity of the fluctuations $\sigma$. This is disproportionally true for agents of relatively high weight. It is thus beneficial to prevent these agents from considering arbitrage altogether. The way this can be achieved is by threatening the agents with legal prosecution if they are identified to be involved in arbitrage (such measures have been applied already to the energy market \cite{50_hertz_untersuchung_2019}). Nevertheless, such prosecution may not completely fend off the arbitrageurs: If they expect a large enough profit, the prospect of this financial profit may outweigh the risk of being punished by the relevant authorities. To investigate the response of the agents to the threat of legal prosecution, we give the agents an additional course of action, following ref.~\cite{challet_criticality_2003} to a large extent. The agents will refrain from arbitrage altogether if they do not expect the profit to outweigh the risk of being prosecuted.

So far we have assumed that agents always decide to play either $a_i = 1$ or $a_i = -1$. This means that the agents never refrain from arbitrage. If the agents are eager to make a profit this is natural, as $a_i = 1$ makes a profit (up to $\mathcal{O}(1/N)$) whenever $I > R(A)$ and $a_i = -1$ makes a profit (up to $\mathcal{O}(1/N)$) whenever $I < R(A)$ (neglecting the correlations between $\langle a_i \rangle$ and $\langle A \rangle$). Thus, a given forecast of $A$ would prescribe an agent to decide $a_i = \pm 1$. However, in reality agents are not willing to perform arbitrage for arbitrarily small profits: For one, since the behaviour is illegal, they will only take their chances if they expect the profit to be larger than the risk derived from being noticed by the authorities. To include this in the minority dynamics, we give each agent $i$ a risk-aversion $\epsilon_i > 0$ (as in \cite{challet_criticality_2003}). The agents can calculate the expected pay-off for a strategy $s$ from the evaluation $U^t_s$: as the evaluation $U^t_s$ represents the total pay-off an agent would have achieved if he would have always used strategy $s$, the pay-off per time-step that he expects from a strategy $s$ is simply equal to $U^t_s/t$. The dynamics of the minority game is then altered as follows:
\begin{itemize}
    \item If an agent $i$ has a strategy with expected pay-off $U^t_s/t$ larger than $\epsilon_i$ he proceeds as usual;
    \item otherwise, he plays $a_i^t = 0$ independently of the signal $\mu^t$.
\end{itemize}
In other words, if an agent expects the pay-off to be worth the risk, he will perform arbitrage; otherwise he will refrain from arbitrage altogether.

Before actually running the minority game, let us first estimate the effects of this modification. Agents will refrain from playing as long as their expected profit is smaller than $\epsilon_i$. The expected profit (up to $\mathcal{O}(1/N)$) is $\langle a \rangle \langle I - R(A + \eta) \rangle$. An agent will thus refrain from playing unless:
\begin{align}
    \langle a_i \rangle \langle I - R(A + \eta) \rangle \geq \epsilon_i \,.
\end{align}
This requires $A$ to be closer to zero (on average) than for the $\epsilon = 0$ case to increase the average pay-off. The only  way the inequality can be satisfied for a substantial fraction of the agents is a situation in which  a number of agents refrain from playing: In this case $A$ becomes closer to $0$. Since $\frac{\dd R(x)}{\dd x} > 0$, this means $\langle I - R(A + \eta) \rangle$ goes up if $A > 0$, while it goes down if $A < 0$, independently of $I>R$ or $I<R$. These cases imply that for most agents $\langle a_i \rangle > 0$ for $A>0$ and $\langle a_i \rangle < 0$ for $A<0$,  therefore increasing profits of most of the active agents. The same effect cannot be achieved by agents simply changing their expected $\langle a_i \rangle$: If it increases (decreases), $A$ also increases (decreases), thereby \textit{lowering} the average profits.

\begin{figure}
    \centering
    \includegraphics[width = 0.5 \textwidth]{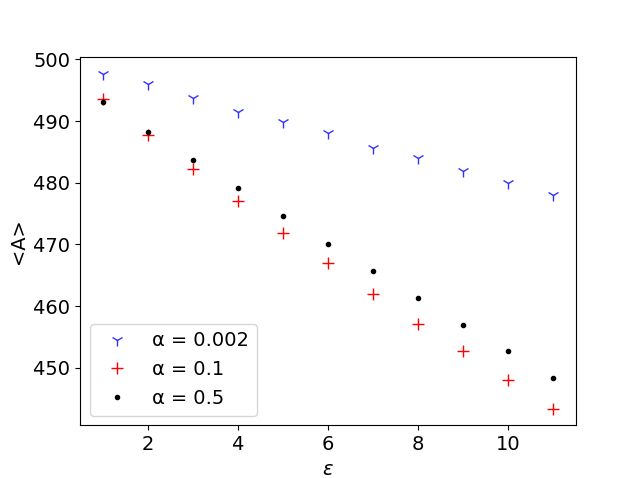}
    \caption{Decreasing expected arbitrage with increasing levels of risk aversion $\epsilon$, for $N = 2000$, $S = 2$, $I = 500$, for different values of $\alpha$. }
    \label{fig: eps_mean}
\end{figure}

To investigate whether this is achieved by the minority game dynamics, we choose $w_i = 1$ for all $i$, $\eta = 0$, linear reserve price function $R(A + \eta) = A$, $I = 500$, $S = 2$, $N = 2000$ and homogeneous $\epsilon$. Strategies are biased as in section~\ref{sec: nonvanishing}, where for a given $s$ and $\mu$,  $a = \pm 1$ occur with probabilities $\frac{1}{2} \pm \frac{1}{2}\frac{A^\ast}{W}$. The resulting $\langle A \rangle$ is shown in Fig.~\ref{fig: eps_mean}, for different values of $\epsilon$ and $\alpha$. It can be seen that the larger the risk-aversion $\epsilon$, the lower the expected amount of arbitrage. Note that $\langle A\rangle$ is not monotonically decreasing with decreasing $\alpha$, but is larger for $\alpha=0.002$. This is likely due to the phase transition that occurs at intermediate $\alpha$.

\begin{figure}
    \centering
    \subfigure{
     \includegraphics[width = 0.45 \textwidth]{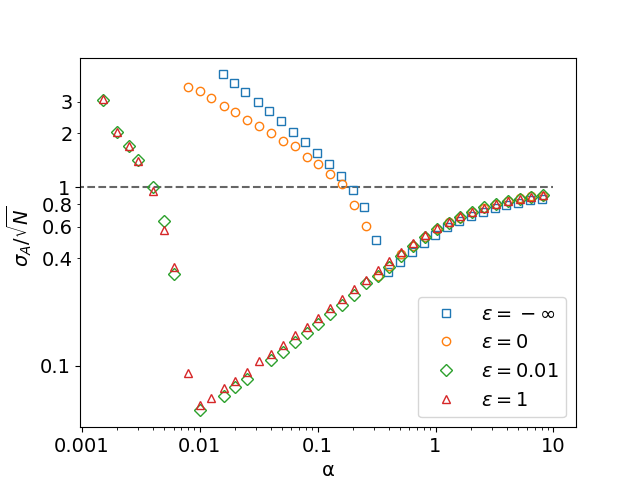}
     \llap{\parbox[b]{17.0cm}{\textbf{\hspace{13.00cm}(a)}\\\rule{0ex}{5.2cm} }}
    }
    \subfigure{
     \includegraphics[width = 0.45 \textwidth]{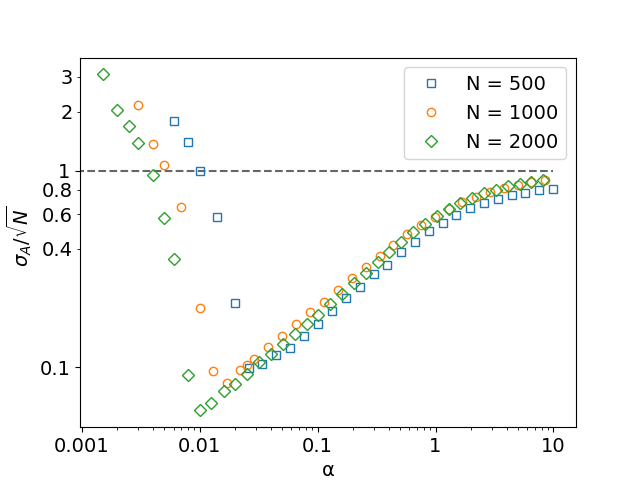}
      \llap{\parbox[b]{17.0cm}{\textbf{\hspace{13.00cm}(b)}\\\rule{0ex}{5.2cm} }}
    }
    \caption{(a) Effect of risk-aversion on the fluctuations of arbitrage, for $N = 2000$, $S = 2$, $I = 500$, as a function of $\alpha$, showing that for positive $\epsilon$ the phase transition occurs at very low values of $\alpha$ (with corresponding low values of $\sigma_A$). (b) Fluctuations of arbitrage for fixed risk-aversion $\epsilon = 1$,  $S = 2$, for different $N$: $N=2000, \, 1000,\,  500$ and $I = 500, \, 250, \, 125$, respectively, showing that results depend not only on $\alpha$, although the high-$\alpha$-phase is almost not affected.}
    \label{fig: eps_sig}
\end{figure}

We investigate whether the structure of the standard version of the minority game (section~\ref{sec: standard}), including the phase transition, remains intact. To this end we run the same version of the risk-averse minority game, and  measure $\sigma_A$ for a wide range of values of $\alpha = P/N$. The results are shown in Fig.~\ref{fig: eps_sig}(a).
The case of $\epsilon = - \infty$ corresponds to the standard minority game discussed in section~\ref{sec: standard}, whereas $\epsilon = 0$ corresponds to the situation in which the agents are willing to perform arbitrage whenever they expect  to make a profit (that is, expect their pay-off to be positive), no matter how small. The results for these two situations are similar, as shown in Fig.~\ref{fig: eps_sig}(a), leaving the overall structure of the results intact. As soon as $\epsilon > 0$, however, a transition to different behaviour is seen. For high values of $\alpha$ the results remain the same as for $\epsilon = - \infty$. However, the location of the phase transition is different: For $\epsilon > 0$, the phase transition occurs at much lower values of $\alpha$, and the standard deviation $\sigma_A$ reaches very low values.

For positive risk-aversion $\epsilon$ there is thus a range of $\alpha$-values for which $\sigma_A$ reaches very low values, largely decreasing the range for which the system is in the $\alpha < \alpha_c$ phase. As this phase would be associated with strong outliers (see Fig.~\ref{fig: histogram}), even a small risk-aversion $\epsilon$ has a positive effect for decreasing the risk of reserve power exhaustion.

We also investigate the question of whether $\alpha \equiv P/N$ remains the only control parameter determining the behaviour of the system. In Fig.~\ref{fig: eps_sig} (b) we show the same risk-averse game as before, with $\epsilon = 1$, for different values of $N$. If $\alpha$ is truly the only control parameter (as it was for the standard minority game of section~\ref{sec: standard}), the results should be the same for different values of $N$, as long as $\alpha = P/N$ is kept constant. Interestingly, while for the high $\alpha$ the results in Fig.~\ref{fig: eps_sig}(b) show that this is the case (to a good approximation), the phase transition does not occur at the same value of $\alpha$: Rather, it occurs at approximately constant $P \approx 20$. For $N \rightarrow \infty$ this means that $\alpha_c \rightarrow 0$. Thus, for $P$ of the same order of $N$, the system is always in the $\alpha > \alpha_c$ phase. We note that the authors of \cite{challet_criticality_2003}, who  investigate the $I = 0$ case, find analytically  that the phase transition completely disappears when $N \rightarrow \infty$; it would be interesting to study whether this corresponds to the same mechanism.

\begin{figure}
    \centering
    \includegraphics[width = 0.5 \textwidth]{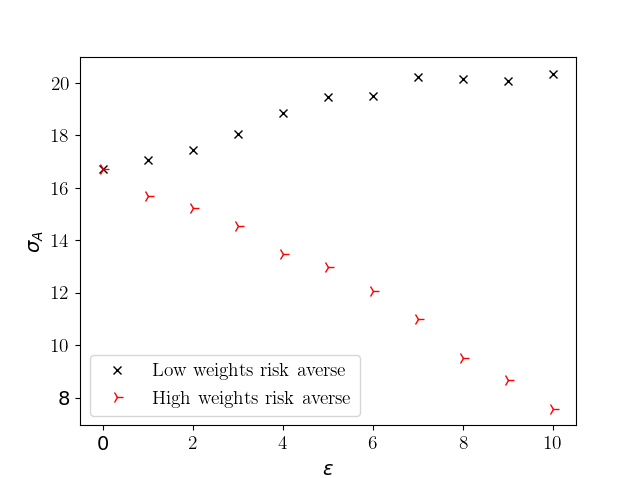}
    \caption{The minority game with heterogeneous risk aversion: Either agents with high weight are assumed to be risk averse while the agents with low weight are not, or vice-versa, leading to different results. Simulated for the realistic weight distribution described in \ref{app: OOM}, $S = 2$, $P = 120$, and $I = 50$.}.
    \label{fig: hetero_eps}
\end{figure}

Finally, we want to investigate another aspect of  risk aversion. In section~\ref{sec: bounds} we found that the 'no anti-coordination' Nash equilibrium suggests that agents with high weight have a disproportionally strong contribution to the fluctuations $\sigma_A$. In order to discourage arbitrage by threatening with penalties, this would imply that the focus should be especially on BRPs that trade in large volumes of power. To investigate this in the context of the learning dynamics of the minority game, we take the realistic weight distribution described in \ref{app: OOM} and split the agents into two groups: Agents with high weight, and agents with low weight, making sure that the total weight in each group is the same. We then give each of the groups a different risk-aversion: Either the group of agents with low weight has $\epsilon = 0$ and the group of agents with high  weight has non-zero $\epsilon$, or vice-versa. We then measure how the overall magnitude of the fluctuations, $\sigma_A$, depends on the strength of the non-zero $\epsilon$, and for which group this non-zero $\epsilon$ is implemented. The results are displayed in Fig.~\ref{fig: hetero_eps}, showing that if agents with high weight have a large value of $\epsilon$, the fluctuations strongly decrease. In the opposite situation, where agents with high weight have $\epsilon = 0$ and agents with low weight have high $\epsilon$, the fluctuations (unexpectedly) increase as compared to no risk aversion.
Naively one may expect that risk aversion of many agents, even agents with low weight, decreases the fluctuations. Thus it is advantageous if agents with high weight behave risk-averse.

\section{Conclusions: Suggestion of Measures for Controlling the Amount of Arbitrage}\label{sec: measures}
As conclusions from the results of the previous sections we suggest some measures for controlling the amount of arbitrage and distinguish between economic incentives via suitable price policies and statutory measures.
\subsection{Economic incentives}
Determining the reserve price by the merit order is essential, as it means the reserve price has a positive first derivative, such that any agent performing arbitrage reduces the arbitrage opportunities for the other agents (the minority mechanism). In addition to this, the most natural way to decrease the incentive for arbitrage is to increase the reserve energy price. Indeed most measures that have actually been implemented have focused on this aspect \cite{50_hertz_untersuchung_2019}.

\subsubsection{Measures via tuning the reserve power price}
\begin{figure}
    \centering
    \includegraphics[width = 0.5 \textwidth]{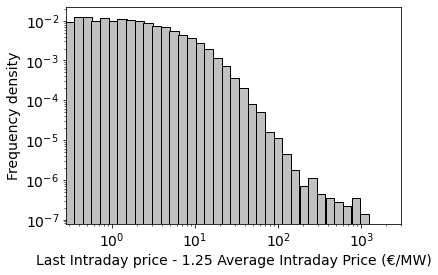}
    \caption{Histogram of the intraday price when the market closes, minus the $1.25$ times the average intraday price  (which, if it is positive, means that there are opportunities for arbitrage). The data shown contains every 15-min interval on the intraday market between 01-01-2017 00:00 and 07-11-2017 24:00, if the difference is positive and the total trading volume on the interval is larger than $500$ MW. Data from \cite{epex_spot}.}
    \label{fig: intraday_difference}
\end{figure}

A significant measure that has been implemented, designed to remove all incentives for harmful arbitrage, is to couple the reserve price to the intraday price \cite{50_hertz_untersuchung_2019, Regelleistung}. We have seen that if the intraday price is higher than the reserve energy price, there is an incentive for the agents to perform the arbitrage corresponding to the decision $a = 1$ (sell energy on the intraday market, feed too little energy into the grid). This increases $A$. On the other hand, if the intraday price is lower than the reserve energy price, there is an incentive for the agents to perform the arbitrage corresponding to $a = - 1$ (buy energy on the intraday market, feed too much energy into the grid). This decreases $A$. From the point of view of the grid stability,  what is desirable is an  imbalance as small as possible. In our framework, this means that $A + \eta$ should be close to zero. Arbitrage increasing or decreasing $A$ is thus harmful if  $A + \eta > 0$ and $A +\eta <0 $, respectively. In these cases the action of the arbitrageur increases the imbalance $A + \eta$, and thus the risk of exhausting all of the reserve power. On the other hand, these actions can also have a positive impact on the security of the grid: If  $A + \eta < 0$ and $A + \eta > 0$, respectively, the arbitrage brings the total imbalance closer to $0$. To make sure that any performed arbitrage is always helpful, a price function must therefore have the following requirements, depending on the sign of the imbalance:
\begin{align} \label{eq: requirements}
\begin{split}
    R(x) > I \quad \text{if} \quad x > 0 \,, \\
    R(x) < I \quad \text{if} \quad x < 0 \,.
\end{split}
\end{align}
After the events of June 2019, this has been pursued by simply implementing a cut-off to $R(x)$ \cite{50_hertz_untersuchung_2019}:
\begin{align}
    R(x) =  \begin{cases}
     - 1.25 \, I_{\text{avg}} & R(x) >  - 1.25 \, I_{\text{avg}} \: \: \text{and} \: \: x < 0 \\
      1.25 \, I_{\text{avg}} & R(x) < 1.25 \, I_{\text{avg}}\: \: \text{and} \: \: x > 0   \\
     R^{\ast}(x) & \text{otherwise} \,,
   \end{cases}
\end{align}
where $R^\ast(x)$ is what the price function would have been without this rule, and $I_{avg}$ is the average price on the intraday market, for a given $15-$minute interval. The interpretation is the following:
\begin{itemize}
    \item If the price function is such that the requirements in Eq. \ref{eq: requirements} hold, there is no intervention.
    \item If the requirements do not hold, the price function is replaced by a constant price for which the requirements do hold.
\end{itemize}
The factor of $1.25$ serves there as a safety margin: $I_{\text{avg}}$ is the average price (for power to be delivered in a given $15-$min interval) on the intraday market. At a specific moment in time the actual intraday price can thus be higher than $I_{\text{avg}}$. Although the cut-off certainly helps to reduce the incentives for arbitrage, for this reason it does not prevent it completely. Figure \ref{fig: intraday_difference} shows that large differences between intraday prices and the average intraday price do occur; in particular, it shows the difference between $1.25\, I_\text{avg}$ and the intraday price at the closing of the market. As the intraday price at closing time often exceeds $1.25 \, I_{\text{avg}}$, sometimes by a large amount, opportunities for arbitrage remain. Changing the cut-off such that the price at closing of the market (rather than the average price) is taken into account may further reduce arbitrage opportunities. In practice the price at closing of the market might be too volatile for this purpose, such that a combination of the average price and of the price at closing may be more suited.

\subsubsection{Measures suggested by the impact of fluctuations}
The measures that have been taken so far have focused on the fact that increasing the reserve power price removes economic incentives for harmful arbitrage. In this paper we have furthermore investigated the implications of the minority mechanism, which gives rise to fluctuations around the expected amount of arbitrage. We found that these fluctuations lead to a significant risk of exhausting a large amount of reserve power (section~\ref{sec: bounds}). So far little attention has been paid to this aspect.

We introduced learning dynamics and found that for a range of parameters (in our model, for $\alpha > \alpha_c \approx 0.34$), the agents learn indirectly to coordinate over time. Thereby they reduce the strength of the fluctuations. In this phase the fluctuations are Gaussian.
On the other hand, we have seen that if the parameters are varied, a phase transition can occur to a phase with very different behaviour. In this phase, fluctuations are strongly non-Gaussian and have large outliers. This behaviour is dangerous for the system, as large fluctuations imply a larger risk of exhausting a significant amount of reserve power.

When investigating the effect of non-linear reserve price function, our results have indicated that fluctuations in the total imbalance change the expected amount of arbitrage (section~\ref{sec: nonlinear}). A high second derivative of the reserve power price (as a function of the total imbalance) decreases the mean amount of arbitrage in the presence of noise or fluctuations in the amount of arbitrage. The stronger the noise or fluctuations, the larger this effect.
In fact, Eq.~(\ref{eq: expected_non_linear}) of the appendix shows that a cut-off in the price function (such as the one discussed earlier in this section) can be expected to have the same effect, where noise (or variance in the amount of arbitrage) decreases the expected profit for the agents, and thereby decreases the expected amount of arbitrage, as the expected price $\langle R(A+\eta)\rangle$ increases.

In view of the impact of fluctuations, one may think of optimizing the design of the market in terms of bounds on the number of participating BRP parties. We have seen a sensible dependence on $N$ (via the parameter $\alpha=P/N$). Our results were restricted to the effect of fluctuations on the arbitrage and therefore not representative for all market activities, but they have indicated that the dependence of $N$ may be non-monotonic  and depend on the distribution of fluctuations which need not be Gaussian. Shortly said, volatility need not increase with only $\sqrt{N}$.

\subsection{Statutory measures}
Strictly speaking, reserve power arbitrage is illegal. Despite this, statutory  measures taken to penalize arbitrageurs can only discourage them to some extent: If they expect a sufficiently large  pay-off, they might take the risk. This is the set-up that we studied in section~\ref{sec: GC}, where we found the following:
\begin{itemize}
    \item Making the agents apprehensive to perform arbitrage decreases the expected amount of arbitrage. The larger the fear of punishment (risk aversion), the lower the amount of arbitrage.
    \item Inspecting the phase structure of the minority game, even a very small fear of penalty significantly decreases the range of $\alpha$ over which the collective dynamics behaves according to the dangerous $\alpha < \alpha_c$ phase. Instead, the critical $\alpha$ is shifted toward smaller values, and the small fluctuations  reveal an effective large degree of anti-coordination  between the agents, dramatically reducing the fluctuations.
    \item Placing an extra focus on statutory prosecution for high-weight agents decreases the fluctuations.
\end{itemize}
As to the overall  conclusion, the threat of statutory prosecution of arbitrageurs has a positive effect on decreasing the risk of reserve power exhaustion, and  placing emphasis on high-weight agents is disproportionally effective at decreasing the fluctuations in the total amount of arbitrage. More surprising is the fact that if the agents experience even a very small risk of legal prosecution, this may already   dramatically decrease the strength of the fluctuations for a certain range of $\alpha$, and thereby decrease the risk of reserve power exhaustion.

\section{Summary and Outlook} \label{sec: summary}
As conclusions from our results we have suggested economic and statutory measures to protect the market from the detrimental effect of arbitrage.
Implementing the mechanism of minority games in the description of the energy market has led to some useful insights which should be relevant also in less stylized and more realistic models of the market. These insights are specific for the physics approach. One is related to the identification of an underlying phase transition (the counterpart in spin glasses is the transition from the replica-symmetry broken phase to the replica-symmetric phase). In our case, varying the dimension of the space of strategies $P$ and/or the number of agents $N$, thereby tuning $\alpha=P/N$, leads from  a phase of unexpectedly strong outliers in the arbitrage to a phase in which the arbitrage is Gaussian distributed (so that outliers are strongly suppressed). Intuitively, such a transition which is responsible for the qualitatively different behavior is not accessible. We have confirmed the occurrence of this phase transition under several extensions of the minimal minority game model. The transition appears rather stable under variations of the input such as the replacement of a uniform distribution of power contributions of the BRP parties to the market by a power-law distribution and a distribution taken from real data.

A second feature which we observed, is well known from complex systems if noise acts in combination with nonlinear dynamics, here the nonlinear reserve price function. The effect of noise can be counterintuitive as confirmed here. More external fluctuations on the energy market (stronger $\eta$) reduce the amount of arbitrage, such that it is possible for the fluctuations to have a positive effect on the balance of the system. The action of noise is even more subtle in the case of colored noise. Together with a nonlinear price function, it then has to be determined from case to case how higher moments of a noise distribution shift the Nash equilibrium toward higher or lower arbitrage, having a beneficial or detrimental effect. These insights on the impact of fluctuations should add upon results from economics which mainly focus on measures directly related to the control of arbitrage rather than on its fluctuations.

For future work it seems interesting to further elaborate on the interaction between various realistic sources of noise and nonlinear price functions. Moreover, since agent-based-modeling should be complemented by an analytical treatment, we think of applying the cavity method as alternative to the replica trick. The goal then would be to minimize the ``implicit information'' in the market by finding the ground state of the related spin-glass problem, here in terms of the strategies used by the agents.

\section*{Declaration of Interest} The authors report no conflicts of interest. The authors alone are responsible for the content and writing of the paper.

\section*{Acknowledgments}
We would like to dedicate this work in memoriam Dietrich Stauffer. One of us (H.M.-O.) is  indebted to Dietrich Stauffer for his constructive support and valuable discussions when changing her field of research from particle physics to network science in 2001. We thank also Martin Palovic (Jacobs University Bremen) for useful discussions from the point of view of an economist and for alluding to reference~\cite{50_hertz_untersuchung_2019}. The authors gratefully acknowledge the support from the German Federal Ministry of Education and Research (BMBF, Grant No. 03EK3055D).

\appendix

\section{Order of magnitude of the fluctuations} \label{app: OOM}
In section~\ref{sec: bounds} we derived as order of magnitude for the ratio of the size of fluctuations and the mean amount of arbitrage the relation:
\begin{align}
    \frac{\sigma_A}{\mu_A} = \frac{ \sqrt{W^2/\mu_A^2 - 1}}{\sqrt{N/X}}\;.
\end{align}
To apply this relation to realistic data, we need to have an estimate of the distribution of weights $w_i$.
As noticed in \cite{50_hertz_untersuchung_2019}, in June 2019 all of the BRPs were imbalanced in the same direction, that is, they did not feed enough power into the grid.
This corresponds to the case where all BRPs play $a = 1$, in which case the distribution of contributions to the imbalance would be equal to the distribution of weights. From \cite{50_hertz_untersuchung_2019} we know that on June 25, 2019 there was a total imbalance of $\approx 5500 \text{ MW}$ (for June 6, 2019 and June 12, 2019, the imbalance was $\approx 5000 \text{ MW}$ and $\approx 7200 \text{ MW}$, respectively). On June 25, 2019, the largest five BRPs caused an imbalance of approximately $2000 \text{ MW}$, the largest ten approximately $2800 \text{ MW}$, the largest twenty approximately $4000$ MW, and in total approximately $5500$ MW. Assuming that within each of these steps the weights are uniformly distributed, and that there are about $120$ BRPs in total (the precise number does not matter a lot, because the contribution of weights which are much smaller than the average contribute a small amount to $N/X$) gives the following distribution in units of $\text{MW}$:
\begin{align}
    w_i =  \begin{cases}
      400 & i = 1, \dots, 5 \\
     160 & i = 6, \dots, 10 \\
      120 & i = 11, \dots, 20 \\
      15 & i = 21, \dots, 120 \,.
   \end{cases}
\end{align}
This leads to $N/X \approx 23.3 $. Estimating $W/\mu_A \approx \sqrt{5} \approx 2.2$ (clearly this depends on $\mu_A$, so this should be taken as an order of magnitude) and combining these results gives $\frac{\sigma_A}{\mu_A} \approx 0.4$. The same approach applied to the data of June 6, 2019 and June 12, 2019 gives  $N/X \approx 18.9$ and $N/X \approx 18.5$, respectively, such that these numbers give  little  higher estimates of $\frac{\sigma_A}{\mu_A}$, around $0.5$. As mentioned before, these estimates should merely be taken as an order of magnitude, but they indicate that the fluctuations are about half of the size of the average arbitrage.

\section{Characteristics of the pay-off function} \label{app: reserve_deriv}
Reserve power is activated according to the merit order, from cheap to expensive. As the total amount of activated reserve power, $x$, increases, the marginal price per unit of reserve power, $\pr(x)$, thus increases as well. The total costs of an amount of reserve power $x$ are given by $\int_0^\mathrm{x} \pr(x') \dd x'$. The average price per unit of reserve power is thus given by:
\begin{align}
    R(x) = \frac{\int_0^\mathrm{x} \pr(x') \dd x'}{x} \,.
\end{align}
Note that $R(x) \leq \pr(x)$ if $x > 0$ and $R(x) \geq \pr(x)$ if $x < 0$. It is a non-decreasing function of $x$:
\begin{align}
    \frac{\dd R(x)}{\dd x} &= \frac{\pr(x) - R(x)}{x} \\
    & \geq \frac{\pr(x) - \pr(x)}{x} = 0 \,.
\end{align}
For positive $R(x)$, assuming that $\pr(x)$ and $x$ have the same sign\footnote{The assumption is that both positive and negative reserve power always comes at a cost. This is in general the case, but for $x$ negative and close to $0$ this assumption does not always hold \cite{Regelleistung}.}, the second derivative is positive if $\frac{\dd \pr(x)/ \dd x}{\pr(x)}$ is large enough:
\begin{align}
    \frac{\dd  {}^2 R(x)}{\dd x^2} &= \frac{1}{x} \big( \frac{\dd \, \pr(x)}{\dd x} - 2 \frac{\dd R(x)}{\dd x} \big) \\
    &\geq \frac{1}{x} \big(\frac{\dd \, \pr(x)}{\dd x} - 2 \frac{\pr(x)}{x} \big) \label{eq: second_deriv_ineq} \\
    &= C(x) \Big(\frac{\dd \pr(x)/ \dd x}{\pr(x)} - \frac{2}{x} \Big)\;,
\end{align}
where $C(x) = \frac{\pr(x)}{x}$ is assumed to be positive. Thus if $\frac{\dd \pr(x)}{\dd x} > \frac{2 \pr(x)}{x}$, the second derivative of $R(x)$ is positive (the reverse does not necessarily hold). If $R(x)$ is negative, the inequality in Eq. \ref{eq: second_deriv_ineq} is reversed, and $\frac{\dd \pr(x)}{\dd x} > \frac{- 2 \pr(x)}{x}$ implies negative second derivative instead (note that in this case $C(x) > 0$ implies $\pr(x) < 0$).

\section{Non-linear price functions} \label{app: non_linear_cost}
 We search for conditions, under which statements about the effect of a nonlinear price function on arbitrage are possible. Assuming that $P(A)$ is symmetric around $A = A^\ast$, we have:
\begin{align}
    \langle R(A + \eta) \rangle &\equiv \int_{-\infty}^{\infty} \dd x \, P(A + \eta = x) R(x) \\
    &= \int_{A^\ast}^{\infty}  \dd x \, P(A + \eta = x) R(x) + \int_{- \infty}^{A^\ast}  \dd x \, P(A + \eta = x) R(x) \\
    &= \int_0^\infty \dd \Delta \, P(A + \eta = A^\ast + \Delta) \big[R(A^\ast + \Delta) + R(A^\ast - \Delta) \big] \,, \label{eq: expected_non_linear}
\end{align}
where $\Delta=\vert A + \eta -A^\ast\vert$. We have:
\begin{align} \label{eq: reserve_deriv}
    \frac{\dd}{\dd \Delta} \big[R(A^\ast + \Delta) + R(A^\ast - \Delta) \big] = \big[R'(A^\ast + \Delta) - R'(A^\ast - \Delta) \big] \,.
\end{align}
 We give two sufficient conditions for this derivative to be positive. For a given $\Delta$:
\begin{itemize}
    \item either $R''(x) > 0$ on the interval $A^\ast  - \Delta < x < A^\ast + \Delta$,
    \item or $R'(A^\ast - \Delta) = 0$ and $R'(A^\ast + \Delta) > 0$ (which is true for the cut-off described in section~\ref{sec: measures}).
\end{itemize}
If one of these conditions holds, the derivative in Eq. \ref{eq: reserve_deriv} is positive. Equation \ref{eq: expected_non_linear} is then of the form:
\begin{align}
\langle R(A + \eta) \rangle = \int_0^{\infty} \dd \Delta \, P(A^\ast + \Delta) \, f(\Delta) \,,
\end{align}
with $\frac{ \dd f(\Delta)}{\dd \Delta} > 0 $ for any $\Delta$ where one of the conditions holds. Thus, within this range, shifting the probability mass to higher $\Delta$ values  (that is, further away from the mean $A^{\ast}$) always increases the expectation value of the price as a function of the arbitrage.

\bibliography{library}

\end{document}